\documentclass[twocolumn,letterpaper,10pt,pre,showpacs]{revtex4-1}
\usepackage{epsfig}
\usepackage{graphics}
\usepackage{amsfonts}
\usepackage{amsxtra}
\usepackage{amsmath}
\usepackage{amsthm}
\usepackage{color}
\usepackage[normalem]{ulem}
\usepackage[utf8]{inputenc}

\renewcommand{\vec}[1]{\boldsymbol{#1}}
\newcommand{\hvec}[1]{\hat{\boldsymbol{#1}}}
\newcommand{\mc}[1]{\mathcal{#1}}

\newcommand{\Eq}[1]{Eq.~(\ref{#1})}
\newcommand{\Eqs}[1]{Eqs.~(\ref{#1})}

\newcommand{\Ref}[1]{Ref.~\citenum{#1}}

\newcommand{\Fig}[1]{Fig.~\ref{#1}}
\newcommand{\Figs}[1]{Figs.~\ref{#1}}

\newcommand{\Sec}[1]{Sec.~\ref{#1}}

\def \ie{{\it i.e.,~}}

\newcommand{\pd}[2]{\frac{\partial #1}{\partial #2}}

\newcommand{\pld}[2]{{{\partial} #1}/{{\partial} #2}}

\newcommand{\bv}[1]{\vec{#1}}
\newcommand{\unitvec}[1]{\hat{\vec{#1}}}

\def\grad{\bv{\nabla}}

\def\Umin{U_{\textrm{min}}}
\def\Elow{U_{\textrm{min}}}
\def\F{\mathcal{F}}
\def\Bm{\bv{B}_{\text{m}}}
\def\Bl{\bv{B}_{\text{p}}}
\def\B0{\bv{B}_{\text{b}}}
\def\muB{\mu_{\text{B}}}

\begin{document}

\title{Nonlinear dynamics of antihydrogen in magnetostatic traps: \\ implications for gravitational measurements}
\author{A.I. Zhmoginov, A.E. Charman, J. Fajans, and J.S. Wurtele}
\affiliation{Department of Physics, University of California, Berkeley, California 94720}
\begin{abstract}
	The influence of 
	gravity on antihydrogen dynamics in magnetic traps is studied.
	The advantages and disadvantages of various techniques for measuring the ratio of the gravitational mass to the inertial mass of antihydrogen are discussed.
	Theoretical considerations and numerical simulations indicate that stochasticity may be especially important for some experimental techniques in vertically oriented traps.
\end{abstract}
\date{\today}
\maketitle


\section{Introduction}
\label{sec:general}

\subsection{Background and Motivation}

	Trapping of neutral antihydrogen was first achieved in 2010 by the ALPHA collaboration~\cite{andr:10a} and, by 2011, ALPHA had reported confinement times up to $1000$ s~\cite{andr:11a}.
	Focus is now shifting from proof-of-principle production and confinement toward precision measurements and tests of fundamental physics.

	There are multiple long-term goals motivating antihydrogen research: the first is to search for possible CPT violation by examining the spectra of anti-atoms.
	A first step in this direction was taken in 2012, when ALPHA  measured the frequency of transitions between hyperfine levels to a relative precision of $10^{-3}$~\cite{Amole:2012a}.
	Future work will concentrate on high precision measurement of this hyperfine splitting, and of the two-photon $1\textrm{S}\to 2\textrm{S}$ transition.

	A second goal is to search for violations of the weak equivalence principle --- the equality of the inertial and gravitation mass of any object, independent of its composition or structure.
	Initial experiments with sensitivity to gravitational effects of the earth on neutral antimatter have been conducted \cite{fajans:12,gabr:12}, and others are planned \cite{kell:08,char:11}.
	The ALPHA collaboration has inferred limits on the ratio $\F=\tfrac{M}{m}$ of gravitational mass $M$ to inertial mass $m$ of antihydrogen by carefully comparing the simulated and experimentally determined temporal and spatial distributions of antihydrogen annihilations observed during the slow turn-off of magnetic trap which confined the anti-atoms.
	Values of $\F > 110$ and $\F < -65$ were rejected ~\cite{fajans:12} at a statistical significance level of $95\%$.
	In a 2012 publication on antihydrogen trapping~\cite{gabr:12}, the ATRAP collaboration reported a gravitational measurement that rejects $\F$ values greater than $200$ at a $2\,\sigma$ level.
	Their methodology was mentioned only briefly, but is apparently based on counting annihilation events during \textit{radial} field shutdown in their vertical trap.
	Understanding the possibilities and limitations of these techniques on trapped neutral antimatter provides the motivation for the present work.
	Two other experiments intending to test the weak equivalence principle, AEGIS~\cite{kell:08} and GBAR~\cite{char:11} also rely on the Antiproton Decelerator (AD) at CERN, but will use beams of antihydrogen rather than trapped populations, and hence their operation is beyond the scope of this paper.

	We present here a detailed study of the influence of gravity on the nonlinear classical dynamics of trapped antihydrogen and, in particular, how features of the nonlinear dynamics impact gravitational measurement techniques in vertical traps.
	Horizontal traps will be discussed in more detail elsewhere.
	Analysis is performed in some generality, but specific numerical examples are motivated by what we infer are the methodology and, roughly, the field geometry used in \cite{gabr:12}, as well as the possibility of using a vertically-oriented version of the ALPHA trap.
	Unless we have misunderstood the ATRAP methodology, our simulations and analysis show no effect of gravity at the levels of sensitivity claimed in their measurement, or, indeed, at lower sensitivity.

\subsection{Dynamical Framework} 

	ALPHA~\cite{bert:06} and ATRAP~\cite{gabr:08} trap neutral antihydrogen in a quasi-static magnetic minimum created by three sets of external coils: two mirror coils produce a field $\Bm$ confining the anti-atoms axially, and a multipole coil produces a field $\Bl$ confining them radially.
	Both the mirror and multipole fields will exhibit spatial variation, and may also have time dependence.
	These trapping fields are superimposed on a static, uniform background solenoidal field $\B0 = B_{\text{b}} \unitvec{z}$ (a legacy of the charged-species trapping preceding antihydrogen production), which reduces the effective trap depth but also, felicitously, tends to suppress non-adiabatic spin flips in the neutral anti-atoms near the field minimum.
	The total magnetic field at a position $\bv{r}$ and time $t$ is then given by the vector sum
	\begin{equation}
		\label{eq:total_field}
		\bv{B}(\bv{r},t) = \B0 + \Bl(\bv{r},t) + \Bm(\bv{r},t).
	\end{equation}
	The orientation of $\B0$, which here establishes the longitudinal ($z$) axis, will be assumed to be either vertical or horizontal, \ie parallel to or perpendicular to the Earth's local gravitational acceleration $\bv{g}$.

	The present analysis presupposes a number of other simplifying assumptions.
	A semiclassical approach describes the internal states of the anti-atom quantum mechanically while treating the center-of-mass (COM) degrees of freedom classically.
	While we allow for the possibility of matter/antimatter asymmetry in gravitational interactions, such that the anti-atom may have an effective gravitational mass $M$ different from its inertial mass $m$, we shall here presume, consistent with CPT invariance, that each antihydrogen is precisely electrically neutral, so experiences no net external Coulomb nor COM Lorentz forces, but can experience a Zeeman force due to a non-zero expectation value for its magnetic moment, whose magnitude is identical to that of an ordinary hydrogen atom in the corresponding internal quantum state.

	Trapping times are sufficiently long so that we may confine attention to anti-atoms which have relaxed to the orbital ground state~\cite{robicheaux:08}, in which the anti-atom's magnetic moment is dominated by the positron spin, and to low-field-seeking spin states, such that the anti-atoms can be trapped near a magnetic minimum.
	(Those anti-atoms in a high-field seeking state quickly hit the trap wall and are thus not considered here).
	The effects of the magnetic field on the internal state are small since
	\begin{equation}
		\muB B = \tfrac{1}{2} \hbar \Omega_e \ll \alpha^2 m c^2,
	\end{equation} 
	where $\alpha = e^2/\hbar c$ (in Gaussian units) is the fine structure constant, $\hbar = h/2\pi$ is the reduced Planck's constant, $c$ the speed of light \textit{in vacuo}, $m$ is the rest mass of the positron and $e$ is the magnitude of its electric charge, while $\muB = e \hbar/2 m c$ is the Bohr magneton, and $\Omega_e = \Omega_e(\bv{r},t) = e B/m c$ is the local positron gryofrequency, where $B = B(\bv{r},t) = \lvert \bv{B}(\bv{r},t) \rvert$ is the magnitude of the total magnetic field at position $\bv{r}$ and time $t$.
 
	Characteristic antihydrogen translational temperatures $T$ are such that translational motion remains entirely non-relativistic:
	\begin{equation*}
		k_{\text{B}} T \ll m c^2,
	\end{equation*} 
	where here $k_{\text{B}}$ is Boltzmann's constant.
	The temperature should also be sufficiently low so that as the anti-atom translates, the changes experienced in local magnetic field strength remain adiabatic with respect to spin dynamics:
	\begin{equation}
		\sqrt{ \tfrac{ k_{\text{B}} T}{m} } \, \tfrac{\lvert \grad \Omega_e\rvert}{\Omega_e} \ll \Omega_e.
	\end{equation} 
	for spatial positions $\bv{r}$ accessible to trapped anti-atoms.
	A related asumption is that the characteristic radial bounce frequency $\omega_r$ and longitudinal bounce frequency $\omega_z$ are both small compared to $\Omega_e$.
	Under these assumptions, the magnetic moment adiabatically tracks the direction of the field, and the classical COM dynamics is governed by the Hamiltonian
	\begin{equation}
		\label{eq:eqm}
		H = H(\bv{r}, \bv{p}, t) = \tfrac{1}{2m} p^2 + \mu_B B(\vec r,t) + M \,\vec{g} \!\cdot\! \vec{r},
	\end{equation}
	where $p=|\vec{p}|$, $\vec{p} = m \dot{\bv{r}}$ is the momentum of the antihydrogen.
	The $+$ sign in front of $\mu_B$ is appropriate for anti-atoms that are in a low-field seeking hyperfine state --- those that can be stably trapped in a local minimum of $B(\bv{r},t)$.

	The study of antimatter gravitational forces requires examining the antihydrogen dynamics for various assumed values of the ratio $\F=M/m$ and comparing these results with experimental observation.
	The gravitational force modifies the dynamics in ways that depend on trap orientation, on field geometry, on initial conditions, and on the time profile of the trap turn-off.
	In ALPHA, an octupole field provides transverse confinement, and the trap axis is horizontal, perpendicular to $\bv{g}$ \cite{bert:06}.
	ATRAP instead employs a quadrupole for transverse confinement, while the trap axis is vertical, parallel to $\bv{g}$ \cite{gabr:08}.
	 
	In cylindrical coordinates $(r,\phi,z)$ oriented such that the solenoidal field points along $\hvec{z}$, the squared-magnitude of the total field \eqref{eq:total_field} is
	\begin{multline}
		B^2 = B^2(r,\phi,z,t) = B_r^2(r,\phi,z,t)+B_{\phi}^2(r,\phi,z,t) \\ + \left[ B_{\text{m}\,z}(r,z,t)+B_{\text{p}\,z}(r,\phi,z,t) + B_{\text{b}} \right]^2,
	\end{multline}
	where $B_r$ and $B_{\phi}$ are the radial and azimuthal field components, which arise from both the multipole and the mirror coils.
	Clearly, the general form of the trapping potential $\mu_B B$ will depend non-trivially on $(r,\phi,z)$ and possibly on $t$ if field turn-off is modeled.
	However, since the decay of the magnetic fields is very slow, unless otherwise noted, our dynamical analysis will be performed presuming a \textit{frozen} value of the Hamiltonian, and the explicit $t$ dependence in $B = B(\bv{r})$ or therefore $H = H(\bv{r}, \bv{p})$ will generally be suppressed in the notation.

\subsection{Basic Dynamic Considerations: Regular versus Stochastic Trajectories}

	We can gain some basic understanding of the dynamics if we temporarily ignore the radial component of the mirror field and any end effects from the multipole.
	Under these simplifications, an order-$\ell$ multipole field yields $B_r^2+B_\phi^2 \propto r^{2\ell-2}$, with no $\phi$-dependence in field magnitude, \ie $B=B(r,z)$.
	For a vertical trap, the total (magnetic and gravitational) potential $U(\vec{r}) \equiv \mu_B B + M\vec{g}\cdot \vec{r}$ is then axially symmetric and separable, \ie $U(\vec{r})= U_r(r)+U_z(z)$, and the longitudinal antihydrogen motion along $z$ is {\em uncoupled} from the transverse motion in the $(r,\phi)$ plane.
	The trajectories are regular and fully determined by integrating two one-degree-of-freedom Hamiltonian systems, namely
	\begin{subequations}
		\label{hamiltonians}
		\begin{align}
			\label{parallel_H}
			H_{\|}(z, p_z) &= \tfrac{1}{2m} p_z^2 + U_z(z),\\
			\label{perp_H}
			H_{\perp}(r, p_r; p_\phi) &= \tfrac{1}{2m} p_r^2 + \tfrac{1}{2 m r^2} p_\phi^2 + U_r(r),
		\end{align}
	\end{subequations}
	where $p_r$ and $p_z$ are, respectively, the radial and longitudinal components of the momentum, and the azimuthal component $p_{\phi}$ represents the angular momentum along $\hvec{z}$.

	In this situation, there are three dynamical invariants: the perpendicular energy $H_\perp = E_\perp$, the parallel energy $H_\parallel = E_\parallel$, and the angular momentum $p_\phi = m r v_\phi$, where $v_\phi = (\dot{y} x - \dot{x} y)/r$ is the azimuthal velocity.
	Any one trajectory with total energy $E = E_{\|} + E_{\perp}$ will not be ergodic and will not explore the entire energy hypersurface $H(\vec{r},\vec{p}) = E$.
	Consider the consequences of very slowly lowering the transverse confining field, \ie $U_r \rightarrow 0$ as $t \to \infty$.
	With uncoupled motion there will be no correlation between the axial dynamics, where gravity acts, and the transverse dynamics; the motions are uncoupled and non-ergodic.
	Consider an anti-atom that has a kinetic energy below the level of transverse potential barrier, but above the axial potential barrier.
	Such an anti-atom is confined transversely, but may escape the trap axially.
	Nonetheless, it will remain confined if enough of its energy is tied up in transverse motion; because there is no coupling, it would never come to possess sufficient energy to overcome the axial barrier.

	In more realistic geometries, some amount of coupling will be caused by the radial components of the mirror field, by the octupole end effects, and by higher-order multipole contributions.
	If the coupling were sufficient to make the motion fully ergodic, then an antihydrogen with total energy $E$ exceeding the lowest of the axial trapping potentials $\Umin$ would eventually escape.
	Knowing $\Umin$ would then allow one to obtain a bound on $M$ by varying only the radially confining potential.
	Such ergodicity may have been implicitly assumed in the gravitational discussion in \cite{gabr:12}.

	Since the dynamics in a realistic magnetic field formed by mirror and multipole coils are not fully integrable, nor expected to be fully ergodic, numerical simulation may be required.
	For small coupling, large regions of phase space should remain integrable.
	The KAM theorem~\cite{arnold:06} suggests that the majority of resonant tori will survive sufficiently small perturbations and the corresponding trajectories remain quasiperiodic.
	In this case, many anti-atoms would remain trapped even when, ostensibly, they appear to have sufficient energy to escape axially.

	The remainder of this paper is organized as follows.
	In \Sec{sec:analytics}, a perturbation theory is used to study the influence of coupling on the anti-atom dynamics.
	A discussion of numerical issues and detailed simulation results for a vertical trap with a field profile similar to that of the ATRAP experiment are presented in \Sec{sec:simulations}.
	These results indicate weak coupling between the transverse and longitudinal dynamics, which may be typical for other existing atom traps~\cite{pinske:98} as well.
	An alternative approach to measuring antihydrogen gravitational mass in a vertical trap, which involves turning off the mirror fields (which does not require anti-atom ergodicity), is also discussed.
	Our conclusions are given in \Sec{sec:conclusions}.


\section{Analytical description of single anti-atom motion}
\label{sec:analytics}

	Detailed analysis of single anti-atom dynamics in a magnetostatic trap is crucial for understanding antihydrogen losses, laser cooling of trapped antihydrogen, and limitations of different approaches to measuring the gravitational mass of antihydrogen.
	While there is no general solution for the full three-dimensional anti-atom trajectory in arbitrary fields, the analysis can be considerably simplified and some insight provided by the case of a trap with nearly-separable confining potentials.

	To begin, we apply canonical Hamiltonian perturbation theory~\cite{landau:76:mechanics} to analyze single anti-atom motion and then use obtained results in \Sec{sec:vertical} to compare numerical simulations with analytical predictions.


\subsection{Perturbation theory for weakly coupled motion}
\label{sec:pt}

	Consider antihydrogen motion in an almost separable potential $U(r,\phi,z)$, \ie $U(r,\phi,z)=U_r(r)+U_z(z)+\delta U(r,\phi,z)$, with $\delta U$ much smaller in magnitude than $U_r(r)+U_z(z)$.
	After rewriting the original (frozen) Hamiltonian (\ref{eq:eqm}) as
	\begin{multline}
		\label{eq:Hp}
		H(r,\phi,z, p_r, p_\phi,p_z) = \frac{p_r^2}{2m} + \frac{p_\phi^2}{2mr^2} + \frac{p_z^2}{2m} + \\ + U_z(z) + U_r(r) + \delta U(r,\phi,z) = H_0 + \delta U,
	\end{multline}
	the term $\delta U$ can be considered as a perturbation to the integrable system with integrable Hamiltonian $H_0$.
	For trap designs like those to be discussed in \Sec{sec:designs}, the magnitude of the perturbation $\delta U$ may be comparable in relative magnitude to the trapping well depth (reaching $O(1/3)$ for the vertical quadrupole trap).
	However, at least in our examples, this maximal value for $\delta U$ is accessible to only those anti-atoms that are weakly trapped in both radial and axial directions, so the perturbation theory should provide some insight into more typical trajectories.

\begin{figure}
		\centering
		\includegraphics[width=0.48 \textwidth]{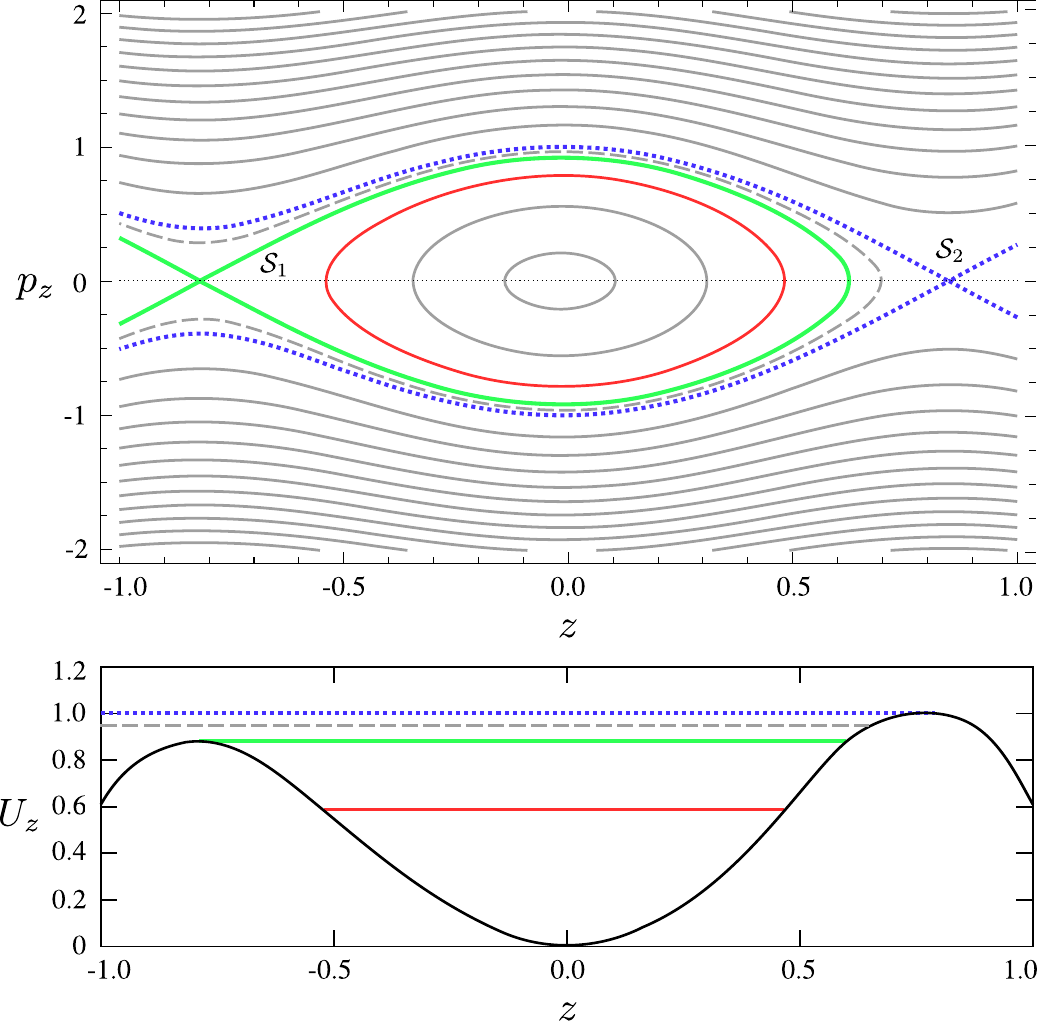}
		\caption
		{
			Phase space portrait of dynamical system governed by $H_\parallel = p_z^2/2m + U_z(z)$ showing two separatrices passing through the lower $\mathcal{S}_1$ (bold solid) and the higher $\mathcal{S}_2$ (dotted) axial potential barriers (top); and a corresponding axial potential profile $U_z(z)$ (bottom).
			The axial coordinate $z$ is normalized to the device half-length $L$, $p_z$ is normalized to $\{2 m \max\limits_{z} [U_z(z)-U_z(0)]\}^{1/2}$, and $U_z$ is normalized to $\max\limits_{z} [U_z(z)-U_z(0)]$.
		}
		\label{fig:phplot}
	\end{figure}

	The first step in applying the perturbation theory to \Eq{eq:Hp} is to find the action-angle variables of the unperturbed Hamiltonian $H_0$.
	The axial motion is uncoupled from the transverse oscillations~\cite{landau:76:mechanics}:
	\begin{equation}
		\label{eq:iz}
		I_z(H_\parallel) = \frac{1}{2\pi} \! \oint p_z\,dz,
	\end{equation}
	where $I_z$ is the axial action, and the integration is performed over a closed trajectory solving $p_z^2/2m + U_z(z) = H_\parallel$.
	The frequency of the axial oscillations is $\omega_z=\dot{\psi}_z=\pld{H_\parallel}{I_z}$, where $\psi_z$ is the angle variable canonically conjugate to $I_z$.
	Assuming that the magnetic field profile $U_z$ is almost quadratic in some vicinity of $z = 0$, $\omega_z(I_z)$ is nearly constant for small $I_z$ (when the anti-atom oscillates near the center of the trap).
	However, when $I_z$ becomes so large that the anti-atom trajectory passes near one of the mirror coils, the frequency $\omega_z(I_z)$ decreases, vanishing eventually at $I_z = I_z^*$, where the anti-atom turning point reaches a local maximum of $U_z(z)$.
	This trajectory corresponds to a separatrix in the $(p_z,z)$ phase space (\Fig{fig:phplot}).

	The radial action $I_r$ can be found similarly.
	First consider a canonical transformation of $H_{\perp} = p_r^2/2m + p_\phi^2/2m r^2 + U_r(r)$ effected by the generating function:
	\begin{equation}
		\Psi(r,\phi,I_r,P_\phi) = \int p_r(r;I_r,P_\phi)\,dr + P_\phi \phi,
	\end{equation}
	where $p_r(r;I_r,P_\phi)$ solves equations of motion for $p_\phi=P_\phi$ and $H_\perp=H_\perp(I_r,P_\phi)$, with $I_r(H_\perp,p_\phi)$ given by
	\begin{equation}
		\label{eq:ir}
		I_r(H_\perp,p_\phi) = \frac{1}{2\pi}\oint p_r(r;H_\perp,p_\phi)\,dr.
	\end{equation}
	After this transformation, $H_\perp$ becomes a function of the new actions $I_r$ and $P_\phi$ and is independent of the new $2\pi$-periodic angles $\psi_r$ and $\psi_\phi = \phi + \Delta$, where
	\begin{equation}
		\Delta(r;I_r,P_\phi) \equiv \! \int\limits^r \pd{p_r}{P_\phi} \, dr.
	\end{equation}
	The canonical angles are generally defined up to an overall constant.
	In the following, we choose $\psi_r=0$ when the anti-atom is closest to the device axis and $\psi_z=0$ when the $z$ coordinate reaches its maximum, and $\Delta(r = 0)=0$.

	It is generally difficult to obtain analytical expressions for the frequencies $\omega_r \equiv \dot{\psi}_r = \pld{H_\perp}{I_r}$ and $\omega_\phi \equiv \dot{\psi}_\phi = \pld{H_\perp}{P_\phi}$.
	Their values are related at $P_\phi=0$, when the anti-atom velocity has a vanishing azimuthal component.
	Introducing the full period of transverse oscillation $T_\perp(I_r)$, one can see that $\phi$ has a period $T_\perp$, while $\psi_r$ has a period $T_{\perp}/2$.
	Then, recalling that $\phi = \psi_\phi - \Delta(\psi_r;I_r,P_\phi)$, one obtains $\psi_\phi(t+T_\perp/2)-\psi_\phi(t)=\phi(t+T_\perp/2)-\phi(t)$ and, therefore, $\omega_\phi T_\perp/2 = \pi$, or $\omega_\phi = \omega_r/2$.


\subsection{Axisymmetric perturbation}

	First, consider the case of a purely axisymmetric perturbation $\delta U(r,z)$.
	The Hamiltonian written in action-angle variables is
	\begin{equation}
	\begin{split}
		\label{eq:Hax}
		H(\vec{I},\vec{\psi}) &= H_0(I_r,I_z) \\&+ \sum\limits_{k=-\infty}^{\infty} \sum\limits_{l=-\infty}^{\infty} \delta U_{k,l}(I_r,I_z) \, e^{i k\psi_r+i l \psi_z},
	\end{split}
	\end{equation}
	where $\delta U_{k,l}$ are the radial/azimuthal Fourier components of $\delta U(\vec{\psi};\vec{I})$.
	The only resonances are, therefore, of the form $k\, \omega_r + l \,\omega_z = 0$.
	This can be rewritten as:
	\begin{equation}
		\label{eq:oeq}
		 k \! \int\limits_{z_1}^{z_2} \! \tfrac{dz}{\sqrt{H_\parallel-U_z(z)}} =
		-l \! \int\limits_{r_1}^{r_2} \! \tfrac{dr}{\sqrt{E-H_\parallel-U_r(r)- \tfrac{P_\phi^2}{2mr^2}}},
	\end{equation}
	where $z_1$, $z_2$, $r_1$, and $r_2$ are the longitudinal and radial turning points, and $H_\parallel$ is a function of $I_z$.
	Note that for the pure quadrupole field with $U_r \sim r^2$, the right-hand side of \Eq{eq:oeq} is independent of $E$, $I_z$, and $P_\phi$, while the left hand side is a function of $I_z$ only.

	Consider a long trap with the radius $R_w$ much smaller than the longitudinal half-length $Z$.
	The ratio $\omega_r/\omega_z$ scales as $Z/R_w$ and the axial scale of the perturbation $\delta U$ will be on the order of the characteristic coil radius $R_c$.
	Since the perturbation maximum is reached at $R \approx R_w$ near the mirrors, the resonance harmonics $\delta U_{k,l}$ and the corresponding resonance widths $\delta I \sim \sqrt{|\delta U_{k,l}|}$ grow with increasing anti-atom energy.
	But even for the highest-energy anti-atoms, $\delta U_{k,l}$ is roughly proportional to $(R_c/L) \max [\delta U] \exp(-k|R_c/R_w|)$ and is small.
	In this case, the radial anti-atom oscillations are adiabatic~\cite{surkov:94,surkov:96}.
	Assuming that most neighboring resonances do not overlap, the system dynamics within resonance islands is expected to be regular, becoming stochastic in small vicinities of the island separatrices only.
	However, since $\omega_z(I_z)$ vanishes at the critical point $I_z^*$ (see \Sec{sec:pt}), there will be an area in the phase space where resonances accumulate and overlap \cite{chirikov:60}, thus forming a stochastic layer in a vicinity of the separatrix at $I_z=I_z^*$~\cite{surkov:94,surkov:96}.

	The systems with axisymmetric perturbations $\delta U(r,z)$ possess another non-generic property which does not survive once azimuthal angle dependence is introduced.
	Namely, for small $\delta U$, \textit{all} antihyrogen trajectories, even stochastic ones, are localized and cannot change their energy by more than a certain finite amount.
	With any amount of angular dependence, this is no longer the case and, in fact, some trajectories in certain time-dependent two-dimensional dynamical systems are known~\cite{arnold:64} to ``diffuse'' indefinitely albeit slowly, reaching any chosen value of action at some sufficiently late moment of time.
	Known as Arnold Diffusion, this phenomenon will be at least partially responsible for slow anti-atom loss from the stationary quadrupole and octupole traps (see \Sec{sec:ergodicity}).


\subsection{Non-axisymmetric perturbation}
\subsubsection{Resonances}

\begin{figure}
		\centering
		\includegraphics[width=0.48 \textwidth]{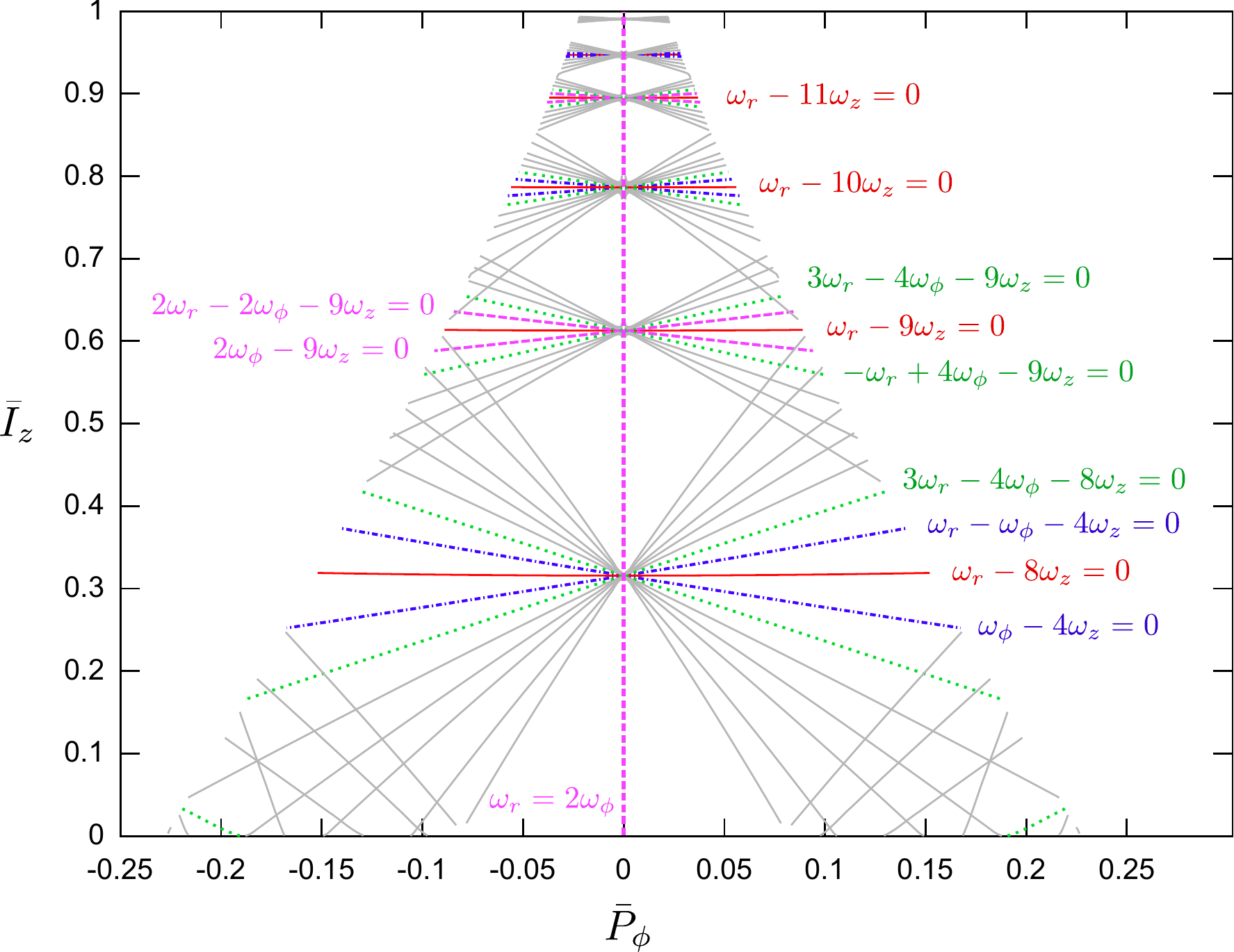}
		\caption
		{
			(Color online)
			Resonances $k \omega_r + l \omega_z + s\omega_\phi=0$ in the $(\bar{I}_z,\bar{P}_\phi)$ space plotted for $E=390\,\textrm{mK}$ in a vertical quadrupole trap: $|s|=0$ (solid red), $|s|=1$ (dash-dotted blue), $|s|=2$ (dashed magenta), $|s|=4$ (dotted green).
			Only resonances with $|k|<10$, $|l|<12$ and $|s|<12$ are shown.
			The action $I_z$ is normalized to the action of the separatrix trajectory corresponding to $\bar{I}_z=1$, and $\bar{P}_\phi = P_\phi/(m R_w v)$, where $v=\sqrt{2 E / m}$ and $R_w$ is the wall radius.
			Note that for a perturbation with $n=4$, all $|s|=1$, $|s|=2$ and $|s|=4$ resonances affect anti-atom dynamics.
		}
		\label{fig:res}
	\end{figure}

	Consider next an angle-dependent perturbation of the form $\delta U(r,\phi,z) = \delta V(r,z)\cos n\phi$ for some fixed integer $n$.
	The Hamiltonian is now:
	\begin{multline}
		\label{eq:Hnax}
		H(\vec{I},\vec{\psi}) = H_0(I_r,I_z) \\ + \tfrac{1}{2}\sum\limits_{k,l} \delta V_{k,l}(I_r,I_z) \left( e^{in\phi}+e^{-in\phi}\right) e^{i k\psi_r+i l\psi_z},
	\end{multline}
	where $\delta V_{k,l}$ is the angular Fourier component of $\delta V(\vec{\psi};\vec{I})$.
	After substituting $\phi = \psi_\phi - \Delta(\psi_r;I_r,P_\phi)$,
	\begin{multline}
		H = H_0(I_r,I_z) \\ + \sum\limits_{k,l} \Bigl( \delta W_{k,l,n}(I_r,I_z)\, e^{ik\psi_r+il\psi_z+in\psi_\phi} + \\ + \delta W_{k,l,-n}(I_r,I_z) \,e^{ik\psi_r+il\psi_z-in\psi_\phi} \Bigr),
	\end{multline}
	where $\delta W_{k,l,n}$ is calculated given all the $\delta V_{k,l}$ as well as $\Delta(\psi_r)$.
	Since the angle-dependent harmonic is fixed, the resonance condition becomes:
	\begin{equation}
		\label{eq:rc}
		Q_{k,l,\pm n}(\vec{I}) \equiv k \,\omega_r + l \,\omega_z \pm n \,\omega_\phi = 0.
	\end{equation}
	Since all frequencies in \Eq{eq:rc} are functions of $I_r$, $I_z$ and $P_\phi$, one can find the resonances in action space.
	Fixing the total energy $H_0(I_r,I_z,P_\phi) = E$, the resonance curves can, for example, be plotted in the $(I_z,P_\phi)$ coordinates.
	Let $\mc{R}^E_n$ be the set of such curves in the $(I_z,P_\phi)$ space corresponding to $Q_{k,l,\pm n} = 0$ for some $k,l \in \mathbb{Z}$.
	Such a plot is shown in \Fig{fig:res} for the quadrupole trap design discussed in more detail in \Sec{sec:vertical}.
	Notice that the resonances $k \, \omega_r + l \, \omega_z = 0$ in \Fig{fig:res} are characterized by nearly constant values of $I_z$, due to the fact that both sides of \Eq{eq:oeq} are independent of $P_\phi$ for the pure quadrupole field.

\subsubsection{Resonance widths}

	Characterizing anti-atom dynamics in phase space requires a knowledge of locations {\em and widths} of all important resonances.
	For sufficiently small $\delta V$, the characteristic width of the resonance $Q_{k,l,s}(I_r,I_z,P_\phi)=0$ is defined by the amplitude of resonant oscillations $\Delta I_r \approx |k| \Delta I$, $\Delta I_z \approx |l| \Delta I$ and $\Delta P_\phi \approx |s| \Delta I$, where~\cite{lichtenberg:92}
	\begin{equation}
		\Delta I = 4 \sqrt{\frac{|\delta W_{k,l,s}|}{|\partial_*^2 H_0|}},
	\end{equation}
	and $\partial_* \equiv k \, {\partial}/{\partial I_r} + l \,{\partial}/{\,\partial I_z} + s\, {\partial}/{\,\partial P_\phi}$.

	Although, for a wide class of smooth functions, the widths $\Delta I$ are expected to decrease exponentially with $| k |$, $| l |$, and $| s |$, the calculation of the exact value of $\delta W_{k,l,s}$ is generally quite complex.
	However, it can be simplified for the $Q_{k,l,0} = 0$ and $Q_{h,0,-2h} = 0$ resonances.
	Indeed, $\delta W_{k,l,0}$ for $n = 0$ is given by:
	\begin{equation}
		\label{eq:wd1}
		\delta W_{k,l,0} = \tfrac{1}{2} \, \delta V_{k,l}.
	\end{equation}
	On the other hand, recalling that $Q_{h,0,-2h}$ vanishes at $P_\phi = 0$, one obtains for $n = 2h$:
	\begin{multline}
		\delta W_{h,0,-2h} = \frac{1}{(2\pi)^2} \sum\limits_{k=-\infty}^{\infty}\int\limits_{0}^{2\pi} \int\limits_{0}^{2\pi} \delta V_{k,0} \cos (2h\phi) \\ \cdot e^{ik\psi_r - ih\psi_r + 2ih\psi_\phi}\,d\psi_r d\psi_{\phi}.
	\end{multline}
	After substituting $\phi = \psi_\phi - \Delta$, this becomes:
	\begin{multline}
		\label{eq:wd2}
		\delta W_{h,0,-2h} = \frac{1}{2\pi} \sum\limits_{k=-\infty}^{\infty} \int\limits_{0}^{2\pi} \frac{\delta V_{k,0}}{2} \\ \cdot e^{2ih\Delta + ik\psi_r - ih\psi_r}\,d\psi_r = \delta V_{0,0}/2,
	\end{multline}
	where we have used the fact that $2\Delta \equiv \psi_r \;(\textrm{mod}\,2\pi)$ for $P_\phi=0$.
	In the following section, we use \Eqs{eq:rc}, (\ref{eq:wd1}), and (\ref{eq:wd2}) to find resonances, estimate their widths, and reach some qualitative conclusions about antihydrogen dynamics in the trap.


\section{Numerical simulations}
\label{sec:simulations}

	In this section, numerical simulations of single anti-atom motion, aimed at assessing the ergodicity of anti-atom trajectories and studying the feasibility of gravitational measurement techniques, are discussed.


\subsection{Computational framework}

	Antihydrogen dynamics were simulated using both standard Runge-Kutta and fourth-order symplectic schemes to integrate the COM equations of motion $\dot{\vec{r}}=+\pld{H}{\vec{p}}$ and $\dot{\vec{p}}=-\pld{H}{\vec{r}}$, where the frozen Hamiltonian $H$ is given by \Eq{eq:eqm}.
	The magnetic field profile $B(\vec{r})$ was calculated from the presumed configurations of magnetic coils at some fixed reference time.
	Since determining the magnetic field using the Biot-Savart law or series expansions for each anti-atom at each moment of time would be quite computationally expensive, we pre-calculated $B$ on a fixed lattice and then interpolated $B$ at instantaneous anti-atom positions.
	Given the representation $B(\vec{r}) = \sum_{n=0}^{\infty} B_{2n}(r,z) \cos (2n\phi+\theta_{2n})$, in the configurations of interest, harmonics $B_{n}$ with $n>2$ for quadrupole traps and $n>4$ for octupole traps can be neglected.
	Therefore, instead of storing a three-dimensional array of $B$ values, we calculated $B_0(r,z)$ and $B_{2\backslash 4}(r,z)$ on a two-dimensional lattice.
	The angular harmonics $B_0$ and $B_{2\backslash 4}$ were calculated using a fast Fourier transform of $B(\vec{r})$ on a ring $(r,\phi_i,z)$, where $\phi_i = 2\pi i/N$ with $N=64$.

	Using only bilinear interpolation to find $B_0$ and $B_{2\backslash 4}$ at some intermediate point would be undesirable, since the force acting on each anti-atom is proportional to $\bv{\nabla} B$, which would then be a discontinuous function causing noise and large numerical errors in antihydrogen trajectories.
	Instead, we used a bicubic interpolation \cite{Note1}, which produced a $C^1$-smooth approximation of $B(\vec{r})$.

	\begin{figure}
		\centering
		\includegraphics[width=0.26 \textwidth]{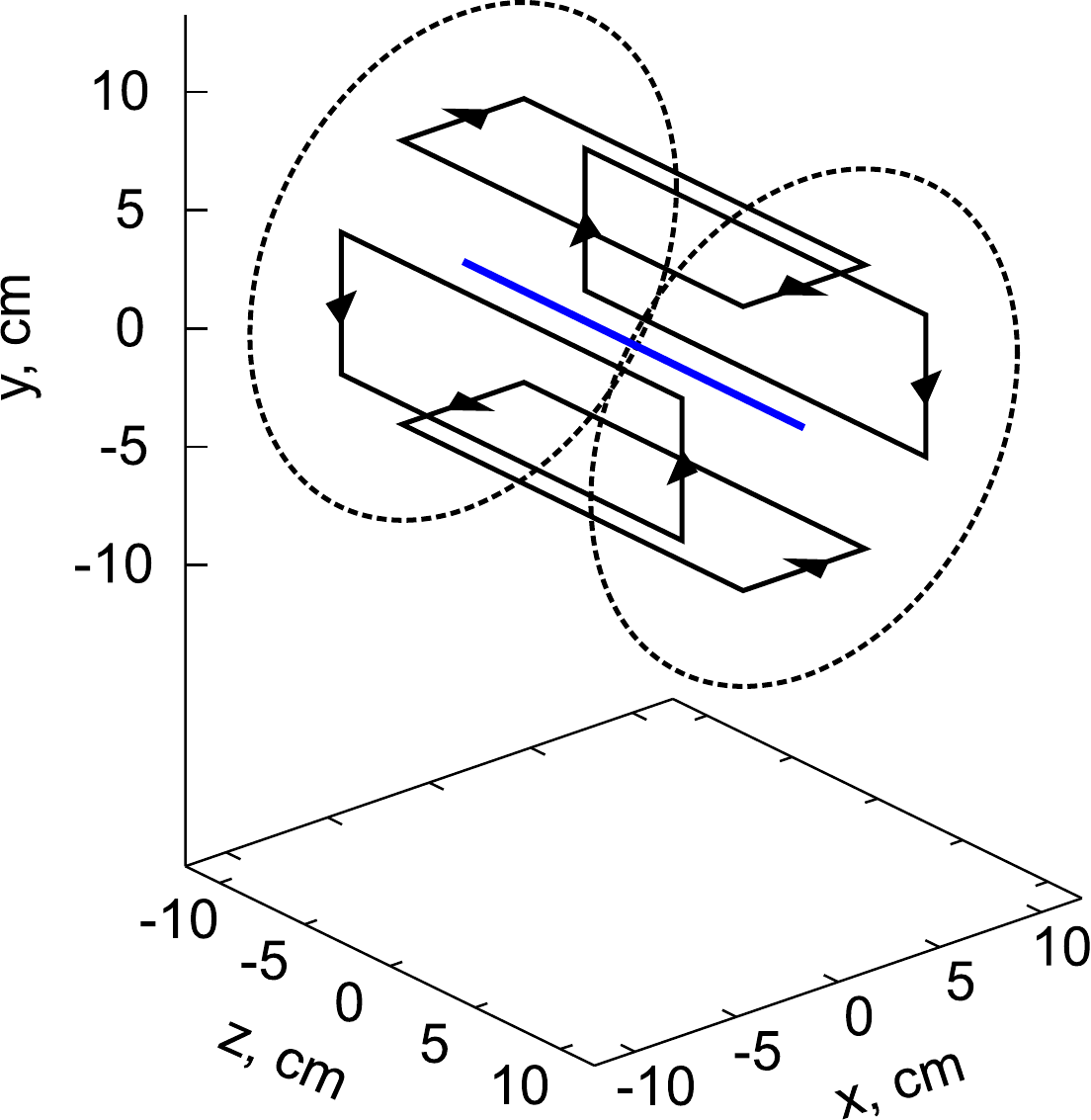}
		\caption
		{
			Set of coils used in our numerical simulations: quadrupole coils (solid) with current directions shown with arrows and mirror coils (dashed).
			The solenoid creating a constant background field is not shown.
		}
		\label{fig:coils}
	\end{figure}


\subsection{Trap geometries}
\label{sec:designs}

	Three device designs were considered: (a) a vertically-oriented quadrupole trap with parameters similar to those of the ATRAP experiment, in which ergodicity and feasibility of gravitational measurements via lowering of the radial confining potential were studied; (b) a vertically-oriented octupole trap otherwise similar to the ALPHA apparatus, which we used to analyze alternative approaches to antihydrogen gravitational mass measurements; and (c) a horizontally-oriented octupole trap similar to the actual ALPHA apparatus (to be discussed elsewhere).

	In all device designs, the background magnetic field $\B0$ was directed along $\hvec{z}$, with magnitude equal to $1\,\text{T}$.
	In a device design motivated by ATRAP \cite{gabr:08,gabr:12} (but not an exact model), two mirror coils of radius $R_p = 10.4\,\textrm{cm}$ were placed at $|z|=Z=10\,\textrm{cm}$ (\Fig{fig:coils}).
	The total current of approximately $265\,\textrm{kA}$ flowing through each mirror coil increased the magnetic field at the trap center to $2.2\,\textrm{T}$, while creating a $375 \,\textrm{mK}$ axial well depth.
	The quadrupole coil was modeled as a combination of 4 rectangular loops with longer sides of length $2Z = 20\,\textrm{cm}$ directed along $\hvec{z}$ and shorter sides of length $R_l\approx 6\,\textrm{cm}$ directed along either $\hvec{x}$ or $\hvec{y}$ (\Fig{fig:coils}).
	Each loop coil located at $| x | = R_l$ or $| y | = R_l$ carried the total current of approximately $360\,\textrm{kA}$.
	As a result, a $375\,\textrm{mK}$ radial well was also created.
	The trap walls, on which the antihydrogen are assumed to immediately annihilate, were chosen to be at $| z |=1.2 \, Z$ and at $r = R_w =1.8 \,\textrm{cm}$.
	In a realistic trap, there are no actual walls at $| z | = 1.2 \, Z$, but all anti-atoms reaching this location will never return to the trapping volume and will annihilate shortly thereafter.
	Performing the angular Fourier decomposition of $B$ inside this volume, one obtains $\lvert B_4(r,z) \rvert < 60\,\textrm{G}$, while $B_0$ is between $2\,\textrm{T}$ and $3\,\textrm{T}$, and $|B_2| \le 0.25\,\textrm{T}$.
	Neglecting octupole and higher-order angular harmonics is therefore justified for this trap.

	In a trap design based on that of ALPHA, the mirror coils located at $| z | = Z = 13.7 \,\textrm{cm}$ created a $670 \,\textrm{mK}$ axial well depth for antihydrogen.
	The octupole coil was modeled as a combination of 8 rectangular loops with longer sides of length $2Z$, located at a distance $R_l = 2.3\,\textrm{cm}$ from the device axis and connected by shorter sides of length $2 R_l \tan(\pi/8)$.
	The magnetic field created by the octupole reached $1.5\,\textrm{T}$ on the trap wall ($R_w = 2.22\,\textrm{cm}$).
	Simulated anti-atoms were assumed to annihilate upon encountering this wall or else when reaching $| z | = L = 15\,\textrm{cm}$.


\subsection{Vertical trap simulation}
\label{sec:vertical}


\subsubsection{Ergodicity of anti-atom trajectories}
\label{sec:ergodicity}

\begin{figure}
		\centering
		\includegraphics[width=0.48 \textwidth]{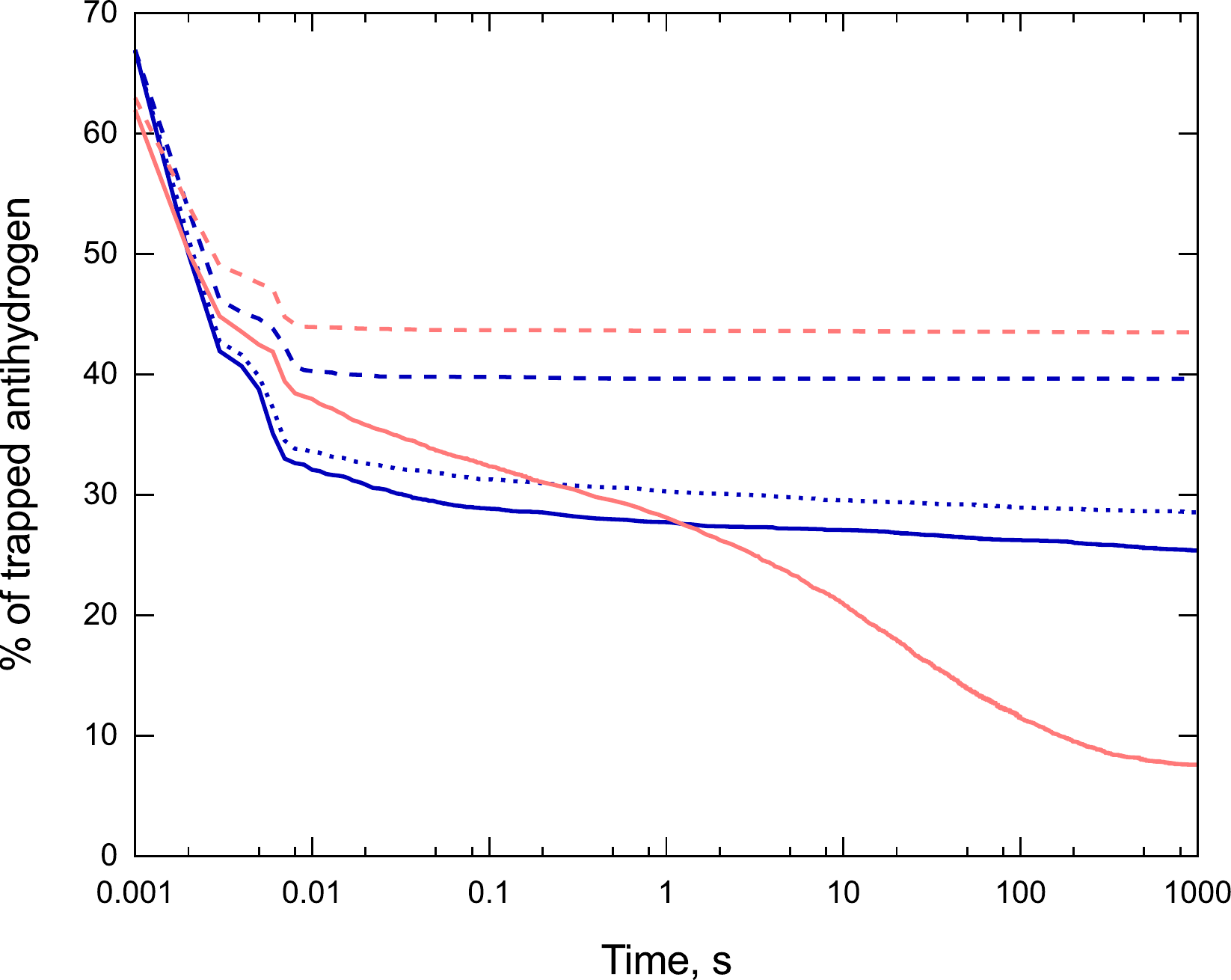}
		\caption
		{
			Fraction $f(t)=n(t)/n(0)$ of anti-atoms remaining trapped in a device as a function of time.
			Simulations were performed for $6000$ anti-atoms with energies within a range $350\,\textrm{mK} \le E \le 550\,\textrm{mK}$ for a trap with a quadrupole coil simulated by $B(r,\phi,z) = B_{q\,0}(r,z) + B_{q\,2}(r,z) \cos 2\phi$ (blue solid) and a trap with an octupole coil simulated by $B(r,\phi,z) = B_{o\,0}(r,z) + B_{o\,4}(r,z) \cos 4\phi$ (red solid).
			Antihydrogens escape from axisymmetric potentials which, although not realizable in multipole traps, share the same angle-averaged profiles with quadrupole and octupole traps were also simulated: a trap with $U(r,z) = \mu B_{q\,0} + Mgz$ (blue dashed) and a trap with $U(r,z) = \mu B_{o\,0} + Mgz$ (red dashed).
			The gravitational to inertial mass ratio $\F=M/m$ was equal to $200$ in all simulations except for one, where anti-atom escape from a quadrupole trap with $B(r,\phi,z) = B_{q\,0}(r,z) + B_{q\,2}(r,z) \cos 2\phi$ assuming $\F=1$ was analyzed (blue dotted).
		}
		\label{fig:log}
	\end{figure}

	As discussed in \Sec{sec:general}, a suggested method of measuring the gravitational mass of trapped antihydrogen by lowering the radial well depth of a vertically-oriented trap and observing annihilations of escaping anti-atoms~\cite{gabr:12} implicitly assumes the ergodicity of anti-atom trajectories.
	The assumption of ergodic trajectories with energy higher than the lowest of the axial trapping potentials, $\Elow\approx 350\,\textrm{mK}$, was verified numerically by simulating anti-atom escape from the vertical quadrupole trap described in \Sec{sec:designs}.
	In our simulations, $6 \cdot 10^3$ anti-atoms were initialized in the trap center homogeneously within a cylinder of radius $1\,\textrm{cm}$ and length $2\,\textrm{cm}$.
	The ratio $\F=M/m$ of the anti-atom gravitational mass $M$ to the inertial mass $m$ was chosen to be $200$, in accordance with the limit asserted in \cite{gabr:12}.
	Initial anti-atom velocities were distributed isotropically, and their energies were chosen randomly and homogeneously from a range $\Elow \le E \le 550\,\textrm{mK}$.
	The anti-atom phase space positions were then numerically evolved in time in static fields for $10^3$ seconds, during which a typical anti-atom performed about $2 \cdot 10^5$ axial and more than $7 \cdot 10^5$ transverse oscillations.
	Every anti-atom encountering the device wall was assumed to annihilate immediately, causing the total number of trapped anti-atoms to drop over time.
	Figure~\ref{fig:log} plots the simulated fraction of anti-atoms remaining in the trap as a function of time, $f(t)=n(t)/n(0)$, from $t=0\,\textrm{s}$ to $t=1000\,\textrm{s}$.
	Different numerical integration schemes showed good agreement, and indicated that after $1000\,\textrm{s}$ more than $25\%$ of all anti-atoms remained trapped in the device with only $2.5\%$ of anti-atoms escaping in the last $999$ seconds.
	That is to say, most escaping antihydrogens escape very early --- about $90\%$ anti-atoms which do escape leave the trap within the first $10\,\textrm{ms}$, which is comparable to a single axial bounce time.
	The fact that in our simulations there exist anti-atoms trapped in the system for $1000\,\textrm{s}$ is not consistent with the assumption of ergodicity; instead, it indicates the existence of bounded regular trajectories.

	It is instructive to see the effect of the angle-dependent harmonic in $B(\vec{r})$ on the anti-atom escape rate.
	In one of our simulations, we considered an axisymmetric potential $U(r,z) = \mu B_{q\,0}(r,z) + Mgz$ identical to the quadrupole potential except for an artificially suppressed $\mu B_{q\,2}(r,z)\cos 2\phi$ term (\Fig{fig:log}).
	This potential has the same angle-averaged profile as in the quadrupole field, but it cannot be physically realized.
	In this case, similar to the previous quadrupole simulation, more than $50\%$ of all anti-atoms escaped within the first $10\,\textrm{ms}$.
	However, $f(t)$ at later times was much flatter in the axisymmetric system, suggesting that the angular resonances may be responsible for a slow anti-atom transport in phase space.
	Indeed, such resonances may lead to Arnold diffusion, which slowly empties resonance layers, driving resonant anti-atoms to the walls.

	The resonance effect of the angular harmonics is even more strongly pronounced in fields with higher multipole perturbations.
	To observe such effects, we simulated anti-atom dynamics in the octupole field by changing the total number of loop currents in the vertical quadrupole trap described in \Sec{sec:designs} from 4 to 8 and also increasing the current $I_q$ by approximately $4$ times to create a similar radial potential barrier, while also reducing the transverse coil size from $R$ to $0.6\, R$.
	The survival fraction $f(t)$ obtained for the octupole field $B=B_{o\,0}+B_{o\,4}\cos 4\phi$ and the axisymmetric potential $U(r,z)=\mu B_{o\,0} + Mgz$ are shown in \Fig{fig:log}.
	Although the axisymmetric potential $U(r,z)=\mu B_{o\,0} + Mgz$ cannot actually be realized in a multipole magnetic trap, simulating anti-atom dynamics in it helps to highlight the role of angular perturbations in long-time anti-atom dynamics.
	\begin{figure*}
		\centering
		\includegraphics[width=0.9 \textwidth]{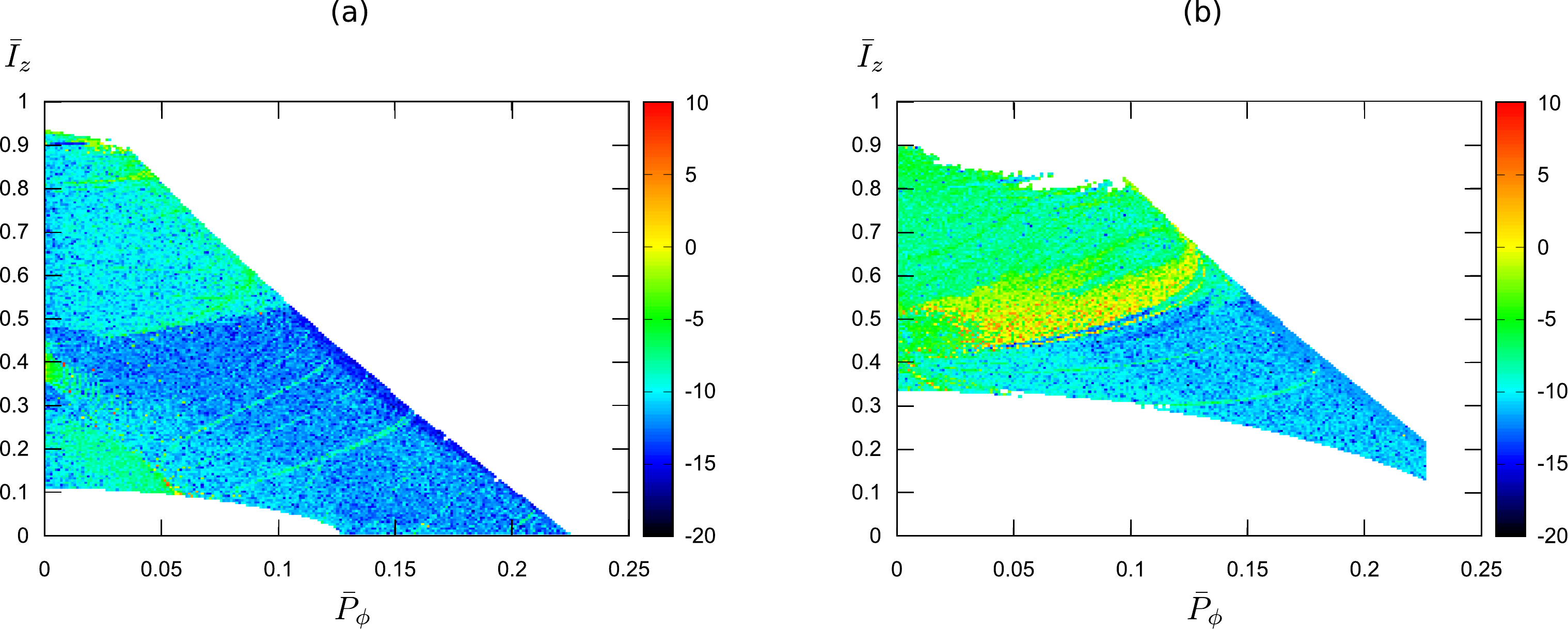}
		\caption
		{
			Frequency map analysis (FMA) diagram~\cite{laskar:90} showing $D(\bar{P}_\phi,\bar{I}_z)=\ln |1-\Omega_B/\Omega_A|$ for $P_\phi(t)$ in a vertical quadrupole trap.
			Primary frequencies $\Omega_A$ and $\Omega_B$ are calculated numerically over two successive $4$ second time intervals $A$ and $B$.
			Blue and green colors correspond to regular quasiperiodic trajectories.
			Values of $\bar{P}_\phi$ and $\bar{I}_z$ together with $\psi_r=\psi_z=\psi_\phi=0$ define the initial conditions for the system trajectory with: (a) $E=390\,\textrm{mK}$ and (b) $E=475\,\textrm{mK}$.
		}
		\label{fig:fmar}
	\end{figure*}

	\begin{figure}
		\centering
		\includegraphics[width=0.4 \textwidth]{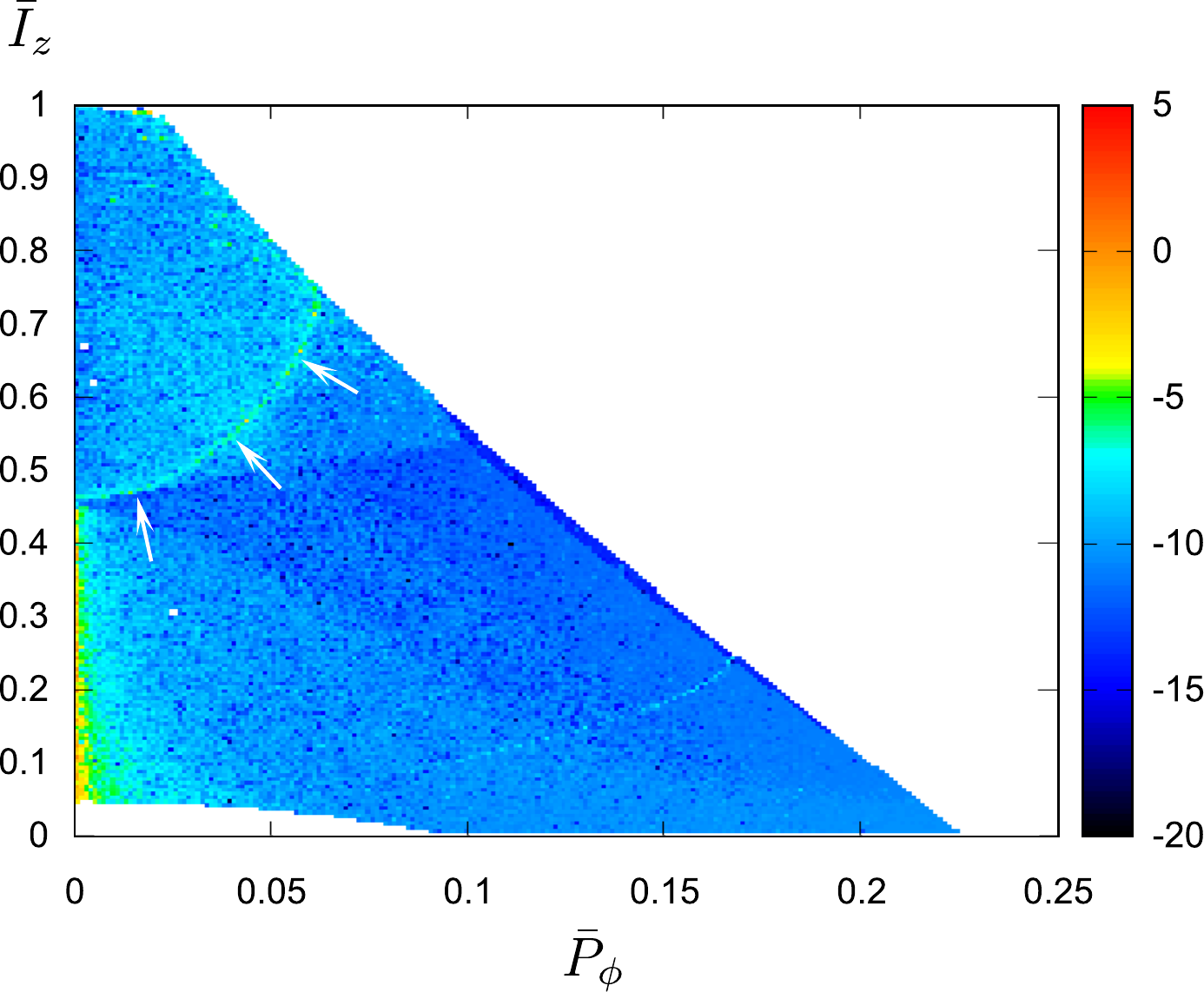}
		\caption
		{
			A FMA map similar to that in \Fig{fig:fmar}a, obtained for $E=390\,\textrm{mK}$ in a system with a trapping potential $U=U_{0} + (U_{2}/4)\cos 2\phi$, where $U_{0}$ and $U_{2}$ are calculated for the vertical quadrupole trap (\Sec{sec:designs}).
			Color palette is altered to highlight the parabolic curve (white arrows) containing strongly nonlinear trajectories.
			It corresponds to the boundary of the resonance island over the stable stationary point (\Fig{fig:avp}).
			The straight line at $\bar{P}_\phi=0$ corresponds to trajectories near the saddle point.
		}
		\label{fig:fma0}
	\end{figure}

	\begin{figure*}
		\centering
		\includegraphics[width=0.8 \textwidth]{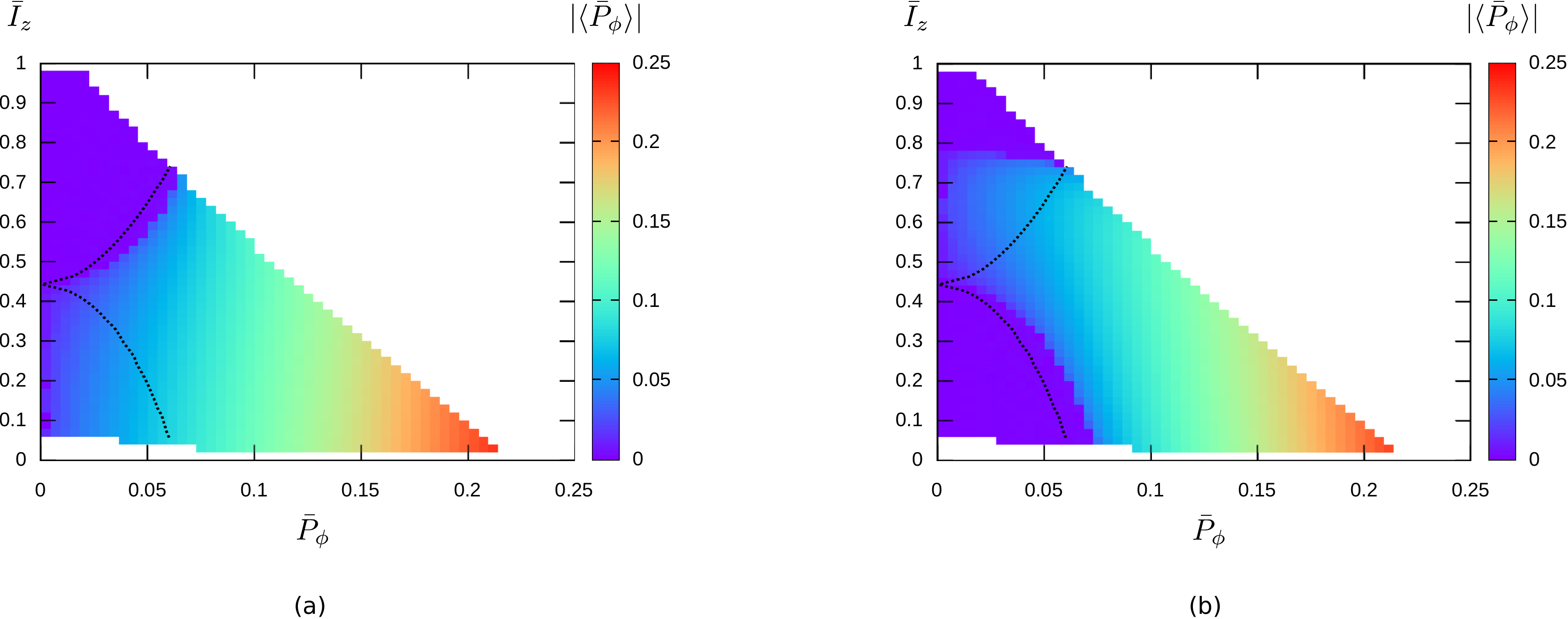}
		\caption
		{
			Dependence of $|\langle \bar{P}_\phi(t) \rangle|(\bar{I}_z,\bar{P}_\phi)$ calculated for $E=390\,\textrm{mK}$ in a system with a trapping potential $U=U_{0} + (U_{2}/4)\cos 2\phi$, where $U_{0}$ and $U_{2}$ are obtained for the vertical quadrupole trap (\Sec{sec:designs}): (a) all anti-atoms are initialized with $\psi_r=\psi_\phi=\psi_z=0$, (b) anti-atoms are initialized with $\psi_\phi=\pi$ and $\psi_r=\psi_z=0$.
			The boundary (dotted line) of the resonance $\omega_r = 2 \omega_\phi$ at $P_\phi=0$ calculated using \Eq{eq:wd2} contains two branches, for which the corresponding one-dimensional phase portraits are shifted by $\pi$ with respect to each other.
			For one of these branches [bottom curve on figure (a)] the initial condition with $\psi_r=\psi_\phi=0$ initializes the system state over the saddle point, hence $\langle P_\phi \rangle$ is non-zero.
			For the other branch, this initial condition places the system state over the stable stationary point, which makes $\langle P_\phi \rangle=0$ within the maximum width of the resonance island.
		}
		\label{fig:avp}
	\end{figure*}


\subsubsection{Comparison with analytical predictions and frequency map analysis}
\label{sec:fma}

	The majority of anti-atoms trapped for more than $10\,\textrm{ms}$ in the quadrupole trap configuration considered above would appear to exhibit regular trajectories.
	This can be explained qualitatively using the formalism outlined in \Sec{sec:analytics}.
	After calculating $I_r$ and $I_z$ numerically, using \Eqs{eq:iz} and (\ref{eq:ir}), the frequencies $\omega_r$, $\omega_\phi$, and $\omega_z$ are obtained by differentiating the unperturbed Hamiltonian expressed as a function of the corresponding actions $\vec{I}$.
	Knowing these canonical frequencies, we identify all resonances for anti-atoms with fixed energy $E$ and plot them in $(I_z,P_\phi)$ space.
	Figure~\ref{fig:res} shows such a plot for a vertically-oriented quadrupole trap (\Sec{sec:designs}) and $E=390\,\textrm{mK}$, with resonances $Q_{k,l,s} \equiv k\, \omega_r + l\, \omega_z + s\, \omega_\phi=0$ and $|k| \le 10$, $|l| \le 12$, and $|s| \le 12$.
	(Without limitation on $k$, $l$, and $s$, the entire plot is covered by a dense set of curves at this resolution).
	The triangular shape of the plot is due to the fact that $H_\perp(I_r,P_\phi)=E-H_\parallel(I_z)$ decreases as $I_z$ increases.

	Not all resonances shown in \Fig{fig:res} influence the dynamics significantly.
	For example, consider a trap with a perturbation possessing only one angular harmonic, \ie $\delta U(r,\phi,z) = V(r,z)\cos n\phi$.
	For $n=0$, corresponding to an axisymmetric perturbation, all resonances have the form $k \, \omega_r + l \, \omega_z=0$.
	For $n=1$, the only resonances affecting anti-atom motion are those $\mc{R}^E_1$ solving $Q_{k, l, \pm 1}=0$ (shown with dash-dotted lines in \Fig{fig:res}).
	The number of resonances increase with a quadrupole field ($n=2$).
	Indeed, every resonance $Q_{k,l,\pm 1}=0$ is also a resonance for $n=2$ since $Q_{2k,2l,\pm 2}=2Q_{k,l,\pm 1}$, \ie $\mc{R}^E_1 \subseteq \mc{R}^E_2$.
	Other resonances $Q_{k,l,\pm 2}=0$ for which either $k$ or $l$ is an odd number, including the resonance $\omega_r=2\omega_\phi$ at $P_\phi=0$ (see \Sec{sec:pt}), are shown in \Fig{fig:res} with yellow dashed lines.
	For the octupole perturbation with $n=4$, the number of resonances increases even further since, again, $\mc{R}^E_1 \subseteq \mc{R}^E_2 \subseteq \mc{R}^E_4$, \ie the set of all resonances $Q_{k,l,\pm 4}=0$ also includes resonances $Q_{k,l,\pm 1}=0$ and $Q_{k,l,\pm 2}=0$.
	This effect may be partially responsible for the presence of a larger fraction of anti-atoms with stochastic trajectories in the octupole traps (\Fig{fig:log}).

	The fraction of anti-atoms affected by a specific resonance depends on its width.
	Using \Eqs{eq:wd1} and (\ref{eq:wd2}), the widths of resonances $Q_{k,l,0}=0$ and $Q_{h,0,-2h}=0$ can be calculated numerically for a vertically-oriented trap with parameters similar to those of ATRAP (\Fig{fig:res}).
	The resonance $Q_{1,0,-2}=0$ is then shown to affect a large fraction of trapped anti-atoms, while $Q_{k,l,0}=0$ resonances have much smaller widths.
	Therefore, since there is no resonance overlap over a large phase space volume, most anti-atom trajectories are expected to be regular.

	The predicted locations of resonances (along with their widths) and the associated stochastic layers can be verified numerically using a variation of the Frequency Map Analysis (FMA) method~\cite{laskar:90}.
	The core idea behind this technique is to test system coordinates like $I_k e^{i\psi_k}$ for quasiperiodicity.
	Treating such a variable as a function of time, one can approximate it as a sum of harmonics $\sum_{k=1}^{N} A_k e^{i\Omega_k t}$ with $|A_{k+1}| \le |A_k|$ and then compare the values of $A_k$ and $\Omega_k$ on different non-intersecting time intervals.
	If the frequencies and amplitudes change considerably along a single anti-atom trajectory, it can be regarded as a sign of stochasticity.
	Figure~\ref{fig:fmar} shows the FMA maps obtained by analyzing $P_\phi(t)$ for initial angles $\psi_\phi = \psi_r = \psi_z = 0$ and energies $E=390\,\textrm{mK}$ and $E=475\,\textrm{mK}$.
	The initial conditions corresponding to chaotic anti-atom motion are shown with yellow and red in \Fig{fig:fmar}.
	This figure and other numerical results obtained for different initial angles suggest that nearly all anti-atoms with energy $E=390\,\textrm{mK}$ are characterized by quasiperiodic trajectories (shown with blue) rather than chaotic motion.
	Higher energy anti-atoms with $E=475\,\textrm{mK}$, however, are more likely to exhibit stochastic dynamics.
	Emergence of stochastic orbits for higher energies can be attributed to the fact that many anti-atoms can now reach regions near the wall ($r=R_w$) at $z\approx Z$, where the angle-dependent perturbation of the trapping potential becomes particularly strong.
	Note, however, that emergence of stochasticity does not necessarily imply rapid anti-atom loss.
	In fact, the majority of anti-atoms with initial states within the bright area in \Fig{fig:fmar}b were shown to stay in the system for at least $100$ seconds.

	Some of the main features of the FMA maps shown in \Fig{fig:fmar} can be related to our analytical predictions.
	Since agreement is better observed for weaker perturbations $\delta U$ (characteristic of traps with smaller radii), consider an artificial system with a trapping potential $U(r,\phi,z)=U_0+(U_2/4)\cos 2\phi$, where $U_0$ and $U_2$ are calculated for a vertical quadrupole trap discussed in \Sec{sec:designs}.
	Calculating locations and widths of resonance islands, all resonances except for $\omega_r=2\omega_\phi$ can be shown to affect only a small region of the system phase space.
	On the other hand, the width of the resonance $\omega_r=2\omega_\phi$ calculated using \Eq{eq:wd2} is sufficiently large to affect almost half of all $390\,\textrm{mK}$ anti-atoms.
	Interestingly, the perturbation harmonic $\delta W_{1,0,2}$ corresponding to this resonance and considered as a function of $I_z$ for $P_\phi=0$ passes through zero at some $I_z=I_*$.
	This means that the phase portrait of the resonance island shifts in phase by $\pi$ after $I_z$ goes through $I_*$.
	As a result, both when $I_z < I_*$, $\psi_\phi=0$ or when $I_z > I_*$, $\psi_\phi = \pi$, the anti-atom is initialized over a saddle point and, thus, all such orbits are not trapped within the resonance, but lie outside of the resonance island.
	On the other hand, for $I_z > I_*$, $\psi_\phi=0$ or $I_z < I_*$, $\psi_\phi=\pi$, the system state is initialized over a stable stationary point, and the corresponding orbit turns out to be trapped for $P_\phi$ smaller than the maximum resonance width.

	These analytical predictions are in agreement with the FMA map shown in \Fig{fig:fma0}.
	Indeed, the bright green round curve in \Fig{fig:fma0} corresponds to the separatrix of the resonance $\omega_r = 2\omega_\phi$, which can also be visualized by plotting the average $P_\phi(t)$ for different anti-atom trajectories (\Fig{fig:avp}).
	If the anti-atom orbit is trapped within this resonance, $\langle P_\phi \rangle = 0$, while for anti-atoms outside of the resonance island, $\langle P_\phi \rangle$ is finite.
	By crossing the separatrix, one would therefore expect to see a jump in $\langle P_\phi \rangle$.
	The analytical predictions for the resonance width and for $I_*$, shown in \Fig{fig:avp} for $\psi_\phi=0$ and $\psi_\phi=\pi$, would seem to agree with the $\langle P_\phi \rangle$ jumping near the actual separatrix.


\subsubsection{Radial barrier shutdown}

\label{sec:vert}

	\begin{figure}
		\centering
		\includegraphics[width=0.48 \textwidth]{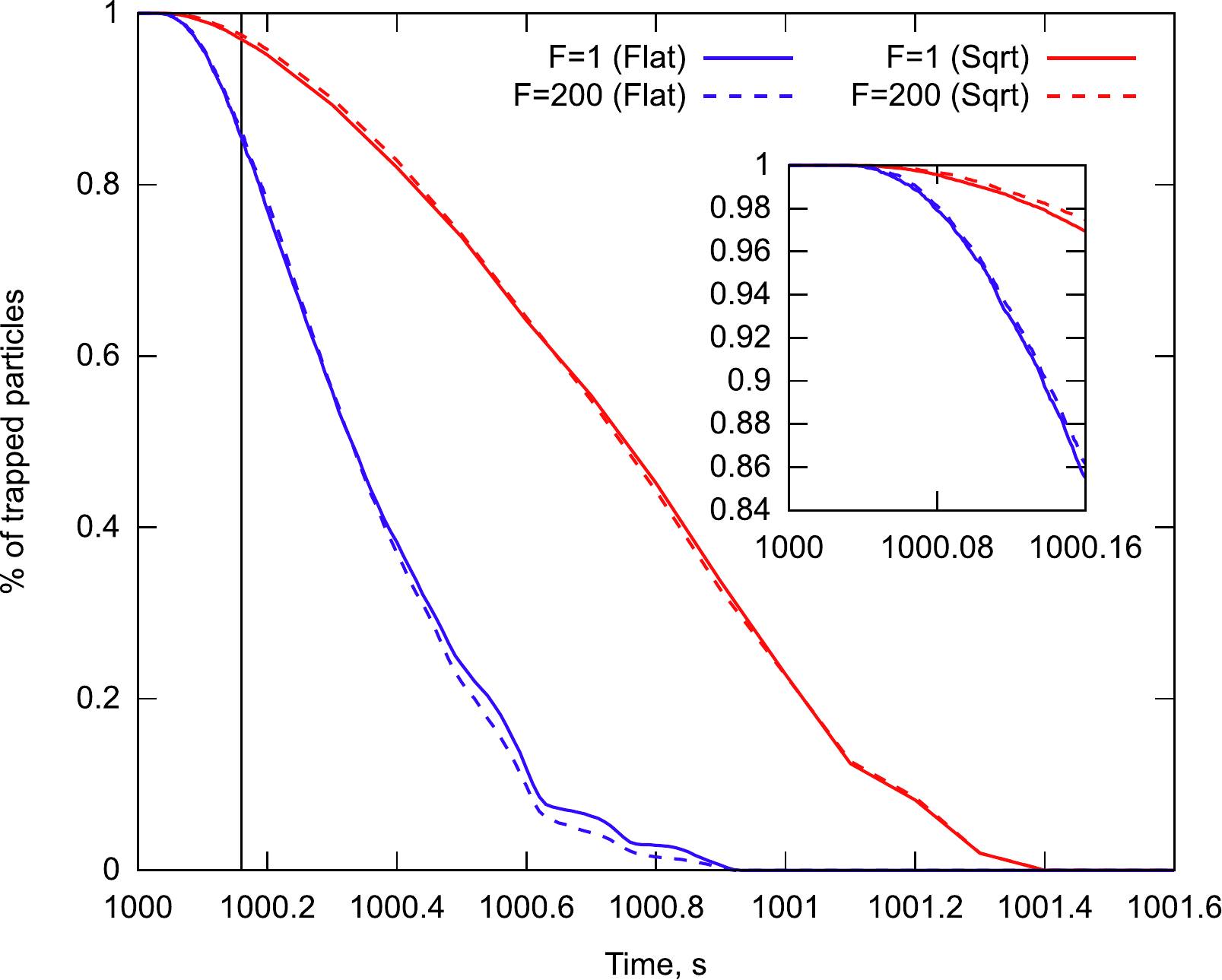}
		\caption
		{
			Survival fraction $f(t)=n(t)/n(t_{\textrm{shut}})$ in the quadrupole trap under the assumption that $\F=1$ (solid) or $\F=200$ (dashed).
			Anti-atoms are initialized with a $\sqrt{E}\,dE$ distribution and $E \le 550\,\textrm{mK}$ (black) and with a flat distribution and $350\,\textrm{mK} \le E \le 550\,\textrm{mK}$ (gray).
			The vertical line corresponds to a time $t_{\text{R}} = 1000.16\,\textrm{s}$ at which the radial trapping potential drops down to $\Elow=350\,\textrm{mK}$.
			The inset shows a zoomed in region for $t<t_{\text{R}}$.
		}
		\label{fig:shut}
	\end{figure}

	In the previous section, based on the numerical simulation of dynamics governed by the frozen Hamiltonian, we inferred that, for our numerical example, a significant fraction of trapped anti-atoms with $E\ge \Elow$ have regular trajectories.
	This makes the assumption of trajectory ergodicity unjustified.
	However, it is still possible that the anti-atom gravitational mass can influence how anti-atoms escape as the radial potential well lowers with the decrease of the quadrupole coil current $I_q$.
	Suppose that the shutdown of the quadrupole coil starts at $t = t_0$.
	If, for a fixed profile $I_q(t)$, the fraction of trapped anti-atoms $f(t)=n(t)/n(t_0)$ is different for different values of $\F$, one can use an experimental measurement of $f(t)$ to infer bounds on the gravitational mass.
	In the following, we compare simulations of $f(t)$ for $\F=1$ and $\F=200$.

	We next identify the multipole field with the field $\bv{B}_{\text{q}}(\bv{r},t)$ created by the quadrupole coils.
	Introducing $\alpha(t) = I_q(t)/I_{q}(t_0)$ so that $\vec{B}_q(\bv{r},t) = \alpha(t) \,\vec{B}_q(\bv{r},t_0)$, the field strength can be written as:
	\begin{multline}
		\label{eq:B}
		|\vec{B}|=\sqrt{B_z^2+B_r^2+B_\phi^2}
			= \Bigl\{\left[B_{bz}+B_{pz}+\alpha B_{qz}(\bv{r},t_0)\right]^2 \\ + \left[B_{pr}+\alpha B_{qr}(\bv{r},t_0)\right]^2 + \alpha^2 B_{q\phi}^2(\bv{r},t_0)\Bigr\}^{1/2}
			\\ = \sqrt{B_A^2 + 2 \alpha G + \alpha^2 B_{q}^2(\bv{r},t_0)},
	\end{multline}
	where $B_A^2=(B_{bz}+B_{pz})^2 + B_{pr}^2$, $G=B_{pr} B_{qr}(\bv{r},t_0) + (B_{bz}+B_{pz}) B_{qz}(\bv{r},t_0)$, and the subscripts $r$, $\phi$ denote the radial and azimuthal components of vectors, respectively.

	Equation~(\ref{eq:B}) was implemented numerically by tabulating zeroth-order and second-order azimuthal harmonics of $B_A^2$, $G$ and $B_{q}^2(\bv{r},t_0)$ independently.
	The simulated antihydrogen ensemble contained 64,000 anti-atoms with an energy distribution $\mathcal{N}(E) \,dE$ scaling like $\sqrt{E} \,dE$ \cite{andr:11a,amol:12,Note2}.
	All anti-atoms were initialized with energy below $550\,\textrm{mK}$, because any anti-atom with higher energy leaves the device within $10\,\textrm{ms}$.
	We compared the loss of anti-atoms due to the quadrupole coil shutdown for $\F=1$ and $\F=200$.
	For the first $t_{0}=1000$ seconds, the quadrupole coil is energized $\alpha(t<t_0)=1$.
	Then, the quadrupole coil is turned off with a characteristic time scale on the order of one second.
	A choice of $\alpha(t)=\exp[-2(t-t_0)^2/(t-t_0+0.8\,\textrm{s})]$ for $t\ge t_{0}$, similar to the reconstructed radial trapping potential shown in Fig.~3b of \Ref{gabr:12}.

	The time-dependence of the fraction of anti-atoms remaining trapped after the initiation of shutdown is shown in \Fig{fig:shut}.
	According to \Fig{fig:shut}, the dependencies $f(t)$ calculated for $\F=1$ and $\F=200$ are virtually identical.
	Introducing the moment of time $t_{\text{R}} \approx 1000.16\,\textrm{s}$ when the radial potential barrier at $z=0$ drops down to $\Umin$, one observes that, while approximately $3\%$ of anti-atoms escape the device prior to $t_{\text{R}}$ in a system with $\F=1$, about $2.5\%$ of anti-atoms escape over the same time interval when $\F=200$.
	Note that if the anti-atom motion were ergodic, no anti-atom de-trapping would be observed until $t= t_{\text{R}}$ for the case where $\F=200$.

	Of course, the fact that in ATRAP, about $10\%$ of all annihilation events were detected before $t=t_{\text{R}}$ \cite{gabr:12} could be attributed to the fact that the actual anti-atom distribution function might differ significantly from $\sqrt{E}\,dE$.
	Additional simulations performed with a flat distribution function, containing only anti-atoms with energies in a range $350\,\textrm{mK} \le E \le 550\,\textrm{mK}$, were shown to be in close agreement with results obtained for a $\mathcal{N}(E) \, dE \propto \sqrt{E}\,dE$ distribution and, in this case, the fraction of anti-atoms escaping before $t=t_{\text{R}}$ reached $7\%$.
	A small deviation of $0.5\%$ between the graphs of $f(t)$ for $t<t_{\text{R}}$ shown in \Fig{fig:shut} could, in principle, be detected in an experiment.
	Note, however, that we infer that only approximately 4 antihydrogen annihilations (with 5 expected cosmic events) were observed in total in \Ref{gabr:12} in the relevant time region between $t=t_0 = 1000\,\textrm{s}$ and $t = t_{\text{R}}=1000.16\,\textrm{s}$.
	This count rate is at least two orders of magnitude lower than that necessary to resolve the differences between the curves shown in \Fig{fig:shut}.

	We infer from these simulations that one cannot establish a limit of $\F<200$ using the technique described above with a vertical quadrupole trap.
	Note that the two distributions studied here are very different, but lead to the same conclusion.

	In the following section we turn our attention an improved technique.


\subsubsection{Axial barrier shutdown}
\label{sec:vpinch}

	\begin{figure}
		\centering
		\includegraphics[width=0.48 \textwidth]{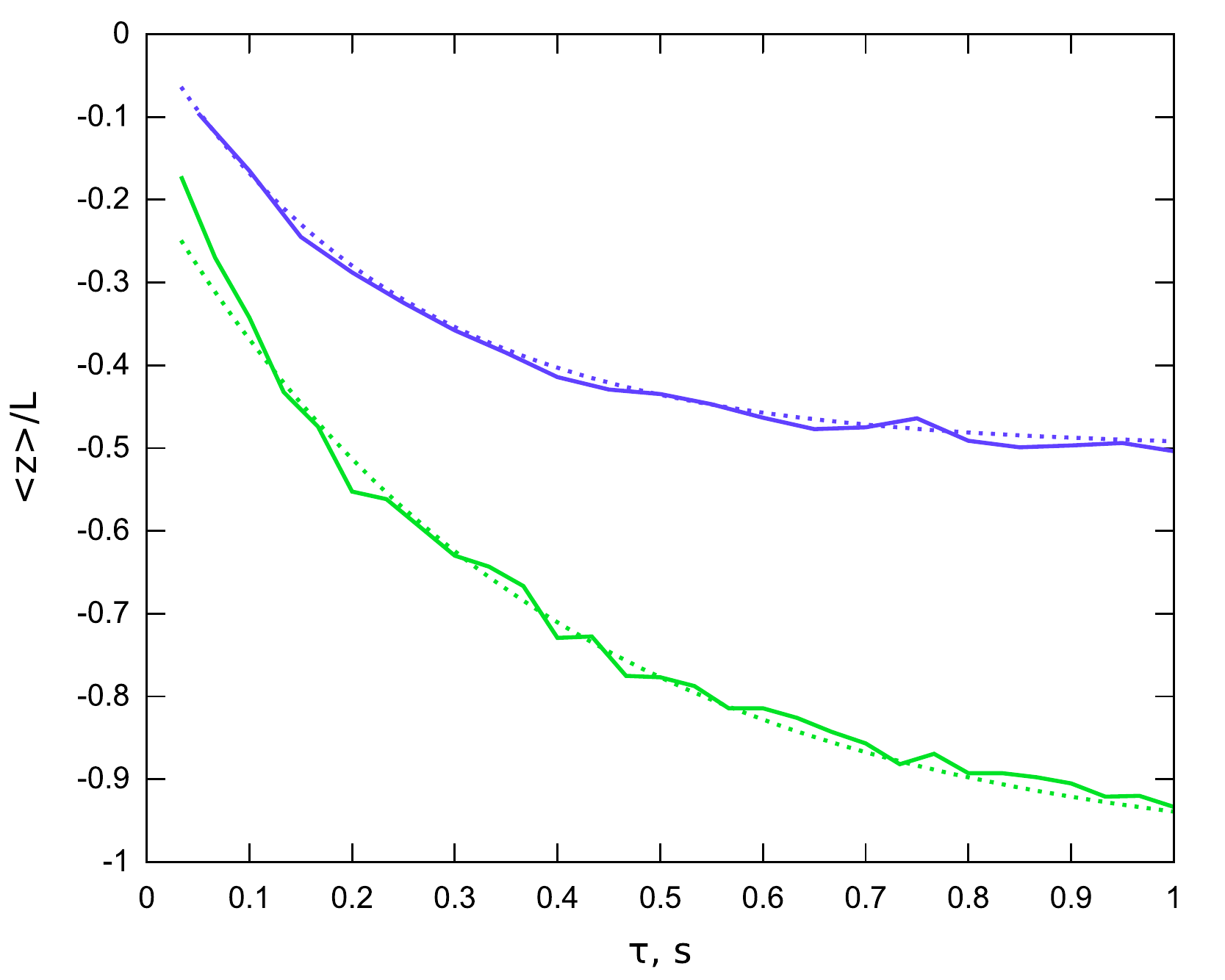}
		\caption
		{
			Dependence of $\langle z \rangle/L$ on $\tau$ for $100\,\textrm{mK}$ antihydrogen atoms trapped in a vertical octupole trap with parameters similar to those of ALPHA (\Sec{sec:designs}) with: (a) artificial separable potential $\bar{U}(r,z)=U(r,0,0)+U(0,0,z)-U(0,0,0)$ (green solid) and (b) realistic trapping potential $U(r,\phi,z)$ (blue solid).
			Two fitted exponential functions $\exp(-2.6 \tau-0.2)-1$ (green dotted) and $0.5(\exp(-4.1 \tau)-1)$ (blue dotted) are shown for reference.
		}
		\label{fig:VA}
	\end{figure}

	A natural alternate approach for measuring the antihydrogen gravitational mass involves lowering the axial trapping barrier in a vertical trap.
	Recall that if $M>0$, the gravitational potential $Mgz$ lowers the trapping potential at the bottom of the trap and raises it at the top, relative to the trap center; for $M<0$, the trapping potential is lowered at the top and raised at the bottom.
	Assuming that currents in both coils are very nearly equal at each moment of time, and that the magnetic field they produce is decreasing in magnitude sufficiently slowly, nearly \textit{all} anti-atoms with $M>0$ will be expected to exit at the bottom of the trap, where the trapping potential is slightly lower (\Fig{fig:phplot}).
	For $M$ negative, antihydrogen would instead preferentially exit the trap at the top.
	Observing the vertical location of antihydrogen annihilations during slow shutdown of mirror coils may, therefore, be a useful experimental technique for quickly assessing the sign of $M$.
	Some preliminary estimates of the required shutdown time and a numerical simulation of such an experiment are discussed below.

	The characteristic {\em adiabatic} time-scale $\tau_*$, on which the trapping potential should be lowered in order to determine the sign of $M$ can, in principle, be estimated by analyzing the axial motion under the Hamiltonian $H_\parallel(p_z,z,t)$.
	Suppose that $M>0$ and consider an anti-atom which is about to cross the inner separatrix $\mc{S}_1$ passing through the saddle point of the lower potential barrier (\Fig{fig:phplot}).
	Let $T(H_\parallel,t)$ be the period of the antihydrogen trajectory calculated for a frozen potential profile $U(z)$, and let $\mc{T}$ be the smallest period of all orbits between two separatrices.
	If $\mc{T}$ is sufficiently large, the anti-atom may cross another separatrix $\mc{S}_2$ passing through the saddle point of the upper potential barrier, after the time $\Delta \tau = (2MgL/U)\tau$, where $U$ is the trap depth and $\tau$ is the actual field shutdown time.
	As a result, the probability for such an anti-atom to leave the device at the top ($p_z>0$) will be approximately equal to the probability of leaving at the bottom ($p_z<0$).
	On the other hand, if $\mc{T} \ll \Delta \tau$, nearly all antihydrogen crossing $\mc{S}_1$ will leave the device at the bottom before reaching $\mc{S}_2$.
	The field shutdown is then adiabatic if it occurs on a time-scale much larger than $\tau_*$ defined by:
	\begin{equation}
		\frac{2MgL}{U} \tau_* = \mc{T}.
	\end{equation}
	The value of $\mc{T}$ can be estimated by recalling that $T(H_\parallel)$ goes to infinity (logarithmically) near both separatrices, and the minimum of $T$ is therefore comparable to
	\begin{equation}
		\mc{T} \sim l_z \sqrt{\frac{2m}{U}} \left[ \ln \left(\frac{4 U}{MgL} \right) + \frac{2L}{l_z} \right],
	\end{equation}
	where $l_z$ is the characteristic scale length of the axial confining potential.
	Assuming that $M=m$, this estimate suggests that for $100\,\textrm{mK}$ antihydrogen with a Gaussian distribution trapped in a device similar to ALPHA, $\tau_*$ is expected to be of an order of the second.
	To verify this conjecture, numerical simulations of antihydrogen escaping from a trap with separable potential $\bar{U}(r,z)=U_r(r) + U_z(z)$, where $U_r(r) = U(r,0,0)$, $U_z(z)=U(0,0,z)-U(0,0,0)$, and $U(r,\phi,z)=\mu B$ is the confining potential of an ALPHA-like apparatus described in \Sec{sec:designs}, were performed.
	Lowering the current in the mirror coils according to $I_{\text{m}}(t) = I_{\text{m} 0} \,e^{-(t-t_0)/\tau}$, we calculated the $z$ coordinates of all simulated annihilation events and plotted their average $\langle z \rangle$ as a function of $\tau$.
	As expected, this average annihilation position $\langle z \rangle(\tau)$, shown in \Fig{fig:VA}, converges to the bottom of the trap $-L$ as $\tau$ goes to infinity.
	The characteristic time-scale of this dependence is on the order of a second, in agreement with the prediction for $\tau_*$.

	\begin{figure}
		\centering
		\includegraphics[width=0.48 \textwidth]{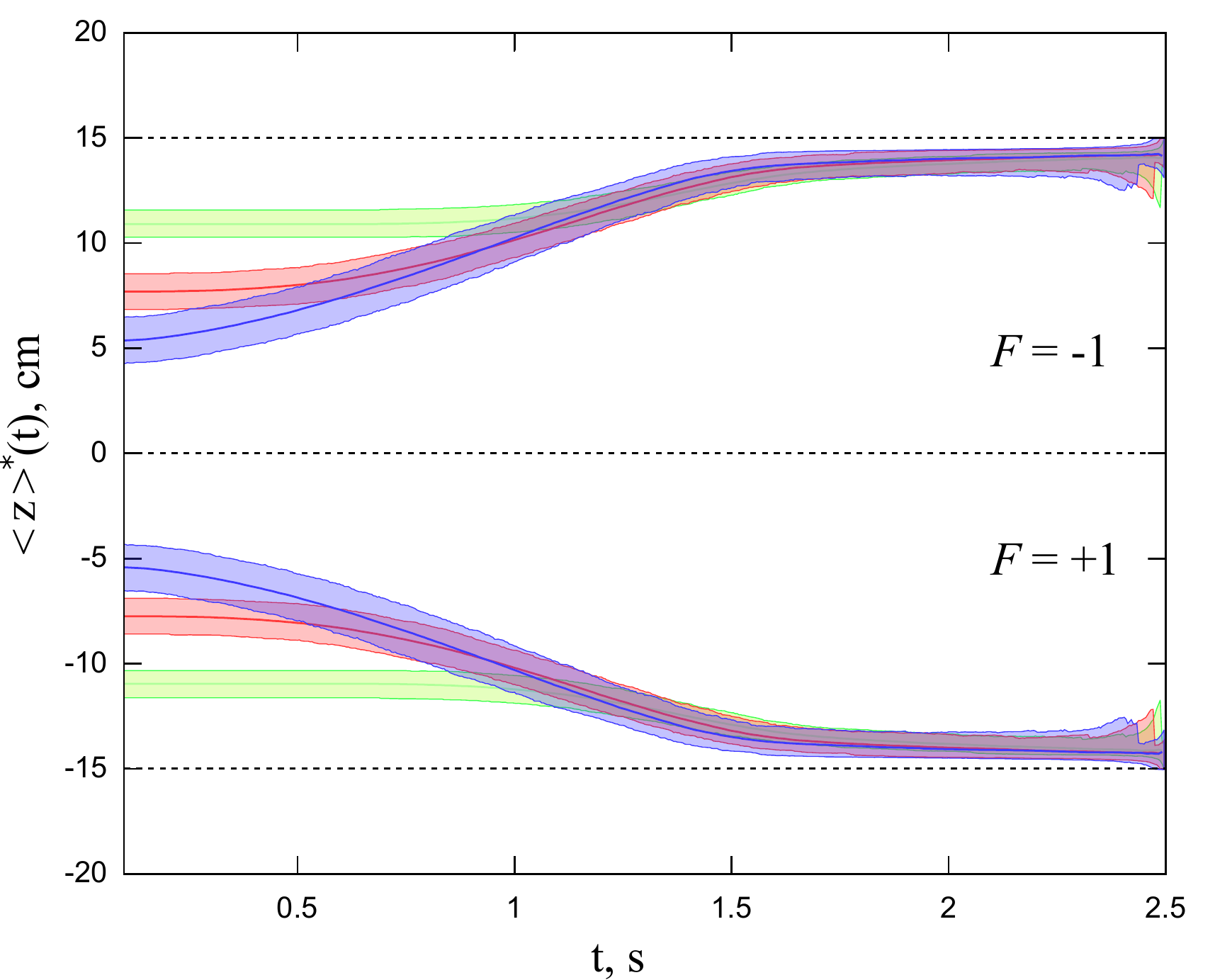}
		\caption
		{
			Reverse cumulative averages $\langle z \rangle^*_\Sigma$ of all $3.2\cdot 10^5$ annihilation events and corresponding confidence regions (blue for $T=300\,\textrm{mK}$, red for $T=100\,\textrm{mK}$ and green for $T=30\,\textrm{mK}$) obtained numerically for anti-atoms with $\F=1$ (bottom) and $\F=-1$ (top) in a vertical trap with $\tau=300\,\textrm{ms}$.
			The assumed particle distribution is $\sqrt{E}\exp(-E/kT)$.
		}
		\label{fig:V1}
	\end{figure}

	\begin{figure}
		\centering
		\includegraphics[width=0.485 \textwidth]{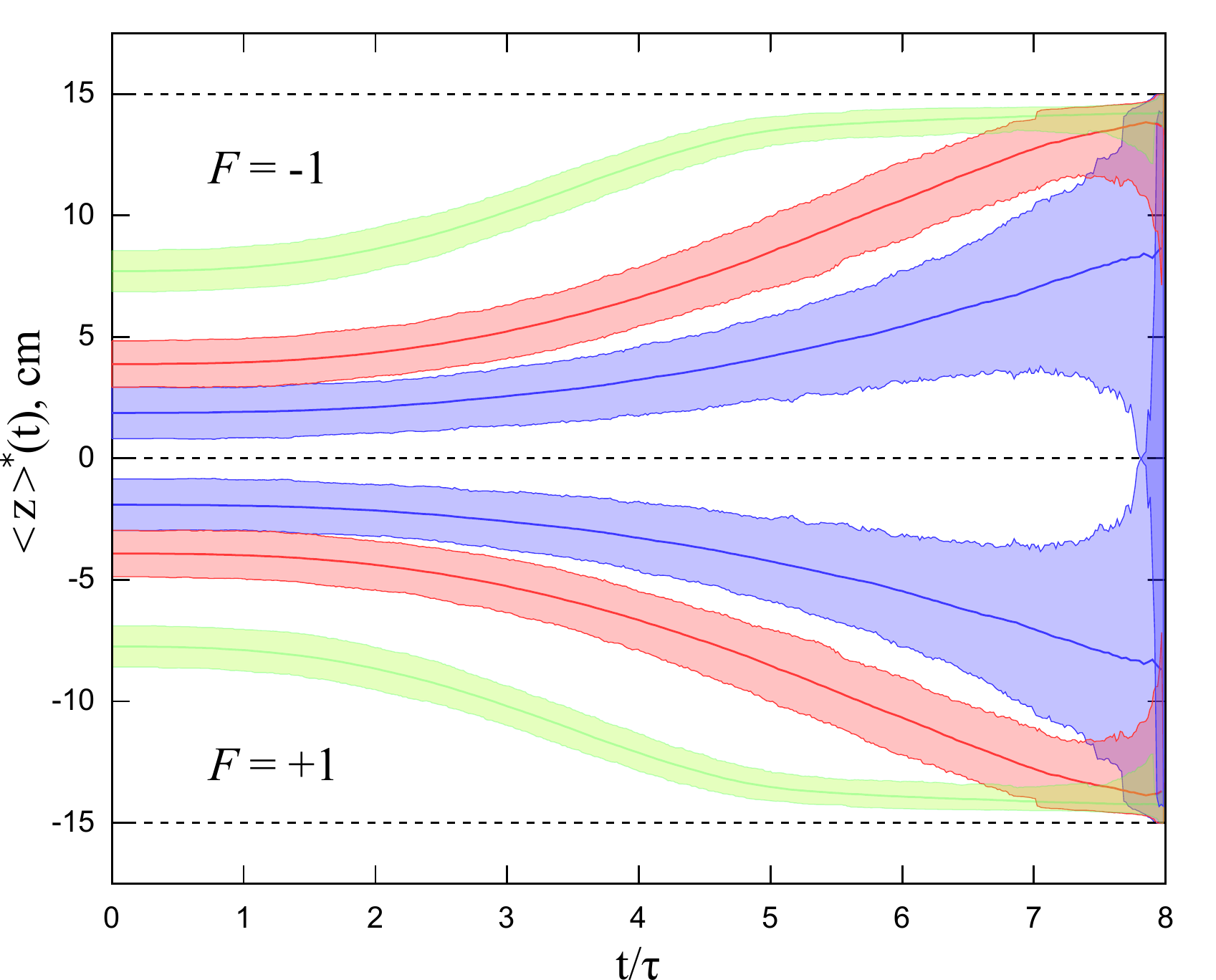}
		\caption
		{
			Reverse cumulative averages $\langle z \rangle^*_\Sigma$ of all $3.2\cdot 10^5$ annihilation events and corresponding confidence regions (blue for $\tau=50\,\textrm{ms}$, red for $\tau=100\,\textrm{ms}$ and green for $\tau=300\,\textrm{ms}$) obtained numerically for $100\,\textrm{mK}$ anti-atoms with $\F=1$ (bottom) and $\F=-1$ (top) in a vertical trap.
			The assumed particle distribution is $\sqrt{E}\exp(-E/kT)$.
		}
		\label{fig:V2}
	\end{figure}

	If implemented, this experimental technique could potentially allow one to distinguish between $\F>\alpha(\tau)$ and $\F<-\alpha(\tau)$ with $\alpha(\tau)\to 0$ as $\tau\to \infty$.
	Choosing a sufficiently large $\tau$, it might even allow one to distinguish $\F \ge 1$ from $\F\le -1$ for even $300\,\textrm{mK}$ antihydrogen atoms.
	Unfortunately, however, this proposed technique would be very sensitive to possible deviations of the actual trapping potential $U(r,\phi,z)$ from its separable approximation $\bar{U}(r,z)$.
	One consequence of the non-separability of $U(r,\phi,z)$ is the emergence of stochastic layers near the separatrices.
	If the layers overlap, anti-atom dynamics within the region confined by $\mc{S}_1$ and $\mc{S}_2$ will be stochastic, and the approximate expression derived for $\mc{T}$ will no longer be valid.
	On the other hand, a small non-separable field component $\delta U = U-\bar{U}$ may perturb low-energy antihydrogen trajectories as the mirror coils are being shut down.
	Indeed, if the oscillations of $I_z$ due to the perturbation $\delta U$ are sufficiently strong and exceed the distance between the separatrices, anti-atoms will cross both of them numerous times.
	As a result, anti-atoms with $M>0$ will retain a finite probability of leaving the device at the top, even if the field shutdown is infinitely slow.

	This effect can be observed by simulating $100\,\textrm{mK}$ antihydrogen escape from a device with a realistic trapping potential $U(r,\phi,z)$.
	Now $\langle z \rangle(\tau)$ does not converge to $-L$ as $\tau\to \infty$, but instead becomes saturated at $\langle z \rangle \approx -7.5\,\textrm{cm}$ (\Fig{fig:VA}).
	Therefore, increasing the shutdown time $\tau$ (beyond about $0.5\,\textrm{s}$ in our case) does not necessarily lead to a substantial decrease of $\alpha(\tau)$ nor to improvement of the antihydrogen mass measurement.

\subsubsection{Accuracy of the gravitational mass measurement}

	As discussed in the previous section, $\langle z \rangle$ of escaping particles measured in a vertical trap with de-energized mirror coils can be very sensitive to the gravitational mass $M$ of the cold antihydrogen.
	This effect could, in principle, be used to determine the value of $M$, or simply check whether $M$ is greater or smaller than zero.
	Here, we determine an accuracy of such a hypothetical test by comparing $\langle z \rangle$ for $\F=-1$ and $\F=1$ assuming that there are only 500 annihilation observations.
	To accomplish this, we follow the procedure discussed in \Ref{fajans:12}, namely we calculate the reverse cumulative averages $\langle z \rangle^*_k(t)$ for $640$ sets $s_k$, each containing $500$ simulated annihilation events.
	This reverse cumulative average is defined as the average $z$ of events occurring after the time $t$, \ie
	\begin{equation}
		\langle z \rangle^*_k (t) \equiv \Biggl( \sum\limits_{i\in s_k,\, t_i < t} z_i \Biggr) \Biggl( \sum\limits_{i\in s_k,\, t_i < t} 1 \Biggr)^{-1}.
	\end{equation}
	Likely statistical fluctuations of $\{\langle z \rangle^*_k\}$ can then be visualized by plotting a confidence region $[z_1(t),z_2(t)]$, chosen in such a way that the intervals $(-\infty,z_1)$ and $(z_2,\infty)$ each contain only $5\%$ of the values of $\{\langle z \rangle^*_k\}$.

	The simulations were performed for a vertical trap with parameters similar to those of the ALPHA apparatus.
	The confidence regions for $\F=\pm 1$, but different values of the antihydrogen temperature and shutdown times are shown in \Figs{fig:V1} and \ref{fig:V2}.
	In these simulations, the time profile of the current in the mirror coils was chosen to be $I_{\text{m}}(t) = I_{\text{m} 0}\, e^{-t/\tau}$ with $\tau = 0.05\,\textrm{s}$, $0.1\,\textrm{s}$, and $0.3\,\textrm{s}$ (we have chosen $t_0 = 0$ for simplicity).
	According to our results, a measurement of $\langle z \rangle$ for $\tau=50\,\textrm{ms}$ and a $\sqrt{E}\exp(-E/kT)\,dE$ particle distribution with $T$ up to at least $600\,\textrm{mK}$ can be used to distinguish between $F=-1$ and $F=+1$ hypotheses with a $95\%$ confidence.
	In other words, a measurement $\langle z \rangle > 0$ ($<0$) is inconsistent with $\F=+1$ ($\F=-1$) hypothesis since this average lies outside of the $95\%$ $\langle z \rangle$-confidence region simulated for $\F=+1$ ($\F=-1$).
	Note that calculating $\langle z \rangle$ of late-time events can further improve the accuracy of the method (\Fig{fig:V1}).
	Simulations performed for a horizontal ALPHA trap suggest that a similar test on the sign of $M$ can be accomplished only for cold plasmas and a large octupole coil shutdown time.
	Fixing the antihydrogen temperature at $30\,\textrm{mK}$, we see that two $95\%$ confidence regions for $\F=\pm 1$ intersect when $\tau < 0.2\,\textrm{s}$.


\section{Conclusions}
\label{sec:conclusions}

	Measuring the ratio of the gravitational to inertial mass in neutral antihydrogen is possible in vertical and horizontal traps, but will require detailed simulations of the nonlinear dynamics of trapped anti-atoms, as these dynamics affect the nature of any signal of the gravitational interaction, and limit the accuracy with which it might be extracted.
	Our study of a vertical quadrupole trap based on the ATRAP experiment shows that the claimed experimental sensitivity is not realized with an experimental methodology inferred from \Ref{gabr:12}.
	Surprisingly, \textit{insufficient} stochasticity can limit schemes to measure the gravitational mass of antimatter.
	In particular, because of a \textit{lack} of ergodicity, radial shutdown in a vertical trap does not appear to offer much sensitivity to $\F = M/m$.
	The coupling of axial and transverse motions and the related notion of stochasticity of typical trajectories in phase space plays especially important roles in other measurement techniques as well.
	In some cases, Arnold diffusion~\cite{arnold:64,lichtenberg:92,rasband:97} and other consequences of stochasticity can limit the precision with which gravitational interactions can be inferred.
	Systematic effects from small field errors and detector misalignments also need to be carefully understood.

\section{Acknowledgements}

This work was supported by the DOE, NSF and LBNL-LDRD.

\end{document}